%% file: theory_paper_v2.tex
\pdfoutput=1 
\documentclass[a4paper,11pt]{article}

\usepackage{jheppub} 
\usepackage{mathrsfs}
\usepackage{amsfonts}
\usepackage{amsmath,amssymb,amsfonts}
\usepackage{slashed}
\usepackage{array}
\usepackage{ulem}
\usepackage{float}
\usepackage{verbatim}
\usepackage{epsfig}
\usepackage{graphicx}
\usepackage{color}
\usepackage[colorlinks=true, linkcolor=blue, citecolor=blue, urlcolor=blue]{hyperref}
\usepackage[dvipsnames]{xcolor}
\usepackage{multirow}
\usepackage{lipsum}
\usepackage{cleveref}
\usepackage{comment}
\usepackage{physics}
\usepackage{mathtools}
\hypersetup{
    colorlinks=true,
    linkcolor=blue,
    filecolor=magenta,
    urlcolor=blue,}

\usepackage[T1]{fontenc} 

\DeclareMathOperator{\sgn}{sgn}
\DeclareMathOperator{\Si}{Si}

\newcommand{\diff}[2]{\operatorname{d}{\hspace{-0.15em}}^{#1}{#2}\hspace{0.15em}}
\newcommand{\calC}{\mathcal{C}}

\title{\Large 
Resummation for Lattice QCD Calculation of Generalized Parton Distributions at Nonzero Skewness
}

\author[a]{Jack Holligan,}
\author[a]{Huey-Wen Lin,}
\author[b]{Rui Zhang,}
\author[b]{and Yong Zhao}
\affiliation[a]{Department of Physics and Astronomy, Michigan State University, East Lansing, MI 48824}
\affiliation[b]{Physics Division, Argonne National Laboratory, Lemont, IL 60439, U.S.A}

\emailAdd{jeholligan@gmail.com}
\emailAdd{hueywen@msu.edu}
\emailAdd{ruizhang@anl.gov}
\emailAdd{yong.zhao@anl.gov}

\preprint{MSUHEP-24-023}

\abstract{Large-momentum effective theory (LaMET) provides an approach to directly calculate the $x$-dependence of generalized parton distributions (GPDs) on a Euclidean lattice through power expansion and a perturbative matching. When a parton's momentum becomes soft, the corresponding logarithms 
in the matching kernel become non-negligible at higher orders of perturbation theory, which requires a resummation. But the resummation for the off-forward matrix elements at nonzero skewness $\xi$ is difficult due to their multi-scale nature. In this work, we demonstrate that these logarithms are important only in the threshold limit, and derive the threshold factorization formula for the quasi-GPDs in LaMET. We then propose an approach to resum all the large logarithms based on the threshold factorization, which is implemented on a GPD model. We demonstrate that the LaMET prediction is reliable for $[-1+x_0,-\xi-x_0]\cup[-\xi+x_0,\xi-x_0]\cup[\xi+x_0,1-x_0]$, where $x_0$ is a cutoff depending on hard parton momenta. Through our numerical tests with the GPD model, we demonstrate that our method is self-consistent and that the inverse matching does not spread the nonperturbative effects or power corrections to the perturbatively calculable regions.}

\begin{document}

\maketitle

\section{Introduction} 

The internal structure of hadrons is a rich topic of study in the field of quantum chromodynamics (QCD).
The phenomenon of confinement means that the constituent quarks, antiquarks and gluons (known collectively as ``partons'') cannot be studied in isolation; their behavior and effects can only be inferred from scattering experiments.
The first function to encapsulate the internal structure of the hadron was the parton distribution function (PDF) which describes the probability density of a parton carrying a specific fraction of the hadron's longitudinal momentum in the limit of the hadron traveling along the lightcone.
A review of these studies can be found, for example, in a Snowmass 2021 whitepaper~\cite{Amoroso:2022eow}.
However, the PDFs only provide a one-dimensional picture of the hadron since they depend solely on the longitudinal momentum.

The generalized parton distributions (GPDs)~\cite{Muller:1994ses,Ji:1996ek,Ji:1996nm,Radyushkin:1996nd} measure not only the parton's longitudinal momentum but also its distribution in the transverse impact parameter space~\cite{Burkardt:2000za,Ralston:2001xs, Diehl:2002he, Burkardt:2002hr}. They provide details on the origin of the mass and spin of the nucleon~\cite{Ji:1996ek}, and their moments and forward limits lead to the gravitational form factors and PDFs. This makes them of great value in the study of hadronic structure. The unpolarized nucleon GPD is comprised of two functions which we denote by $H$ and $E$. In terms of the lightcone matrix elements, the quark GPD is defined as
\begin{align}\label{eq.LightconeGPD}
    F(x,\xi,t) &= \int\frac{\diff{}{z^-}}{4\pi} e^{ixp^+z^-} \left\langle p''\left|\overline{\psi}\left(-\frac{z}{2}\right)\gamma^+ W\left(-\frac{z}{2},\frac{z}{2}\right)
    \psi\left(\frac{z}{2}\right) \right|p'\right\rangle\nonumber\\
    &=\frac{1}{2p^+}\left[H(x,\xi,t)\overline{u}(p'')\gamma^+u(p')+E(x,\xi,t)\overline{u}(p'')\frac{i\sigma^{+\nu}\Delta_{\nu}}{2m}u(p')\right],
\end{align}
where $W(-z/2,z/2)$ is a Wilson line along the lightcone, $\ket{p}$ is a hadron state with 4-momentum $p^{\mu}$, $\psi$ is the fermion field. The momentum transfer $\Delta^\mu=(p''-p')^\mu$, $t\equiv \Delta^2$, and the skewness parameter $\xi=\frac{p''^+ - p'^+}{p''^+ + p'^+}$.
The lightcone coordinates are defined as $z^{\pm}=\frac{1}{\sqrt{2}}(z^0\pm z^3)$ for a hadron moving along the $z^3$ axis.
In the limit $t \to 0$, the $H$ GPD reduces to the PDF, whereas the $E$ GPD is inaccessible for it is multiplied by the momentum transfer vector. Experimental studies of GPDs will be a top target at the future Electron-Ion Collider (EIC)~\cite{Accardi:2012qut,AbdulKhalek:2021gbh,AbdulKhalek:2022hcn,Burkert:2022hjz,Achenbach:2023pba,Abir:2023fpo} through processes such as deeply-virtual Compton scattering (DVCS)~\cite{Ji:1996nm} and meson production (DVMP)~\cite{Radyushkin:1996ru,Collins:1996fb}.

The first-principles lattice QCD calculation of GPDs started with their Mellin moments over two decades ago~\cite{Mathur:1999uf,Gadiyak:2001fe,Gockeler:2003jfa,Hagler:2003jd,LHPC:2003aa,Gockeler:2005cj,QCDSF:2006tkx,LHPC:2007blg,QCDSF:2007ifr,Deka:2013zha,Shanahan:2018pib,Shanahan:2018nnv,Alexandrou:2020sml}. 
However, this method is limited to the lowest few moments due to the worsening signal-to-noise and power-divergent operator mixing.
The proposal of large-momentum effective theory (LaMET) in 2013~\cite{Ji:2013dva,Ji:2014gla,Ji:2020ect} made it possible to directly calculate the $x$ dependence of PDFs and GPDs, which has led to significant progress in this field along with other approaches~\cite{Liu:1993cv,Detmold:2005gg,Detmold:2021uru,Braun:2007wv,Chambers:2017dov,Radyushkin:2017cyf,Ma:2017pxb,CSSMQCDSFUKQCD:2021lkf,Bhattacharya:2024qpp,HadStruc:2024rix,Hannaford-Gunn:2024aix} over the years.
The LaMET approach involves computing spatially separated correlators on the Euclidean lattice and relating the corresponding momentum-space distributions---the quasi distributions---to the lightcone through effective theory expansion and matching~\cite{Ji:2015qla,Xiong:2015nua,Liu:2019urm,Ma:2022ggj,Ma:2022gty,Yao:2022vtp}.
The first GPD to be studied in the LaMET framework was for the pion in Ref.~\cite{Chen:2019lcm} with an unphysical pion mass of $310$ MeV.
The equivalent calculations in the zero-skewness limit were performed by MSULat at the physical pion mass~\cite{Lin:2023gxz,Lin:2023kxn} as well as by the BNL-ANL group at a superfine lattice spacing with unphysical quark masses~\cite{Ding:2024hkz}.
The unpolarized and helicity GPDs for the nucleon were later studied in Refs.~\cite{Alexandrou:2020zbe,Lin:2020rxa,Lin:2021brq,Bhattacharya:2022aob,Bhattacharya:2023jsc} and the transversity GPD was calculated in Ref.~\cite{Alexandrou:2021bbo}. The twist-three GPDs were also explored by the ETMC in Ref.~\cite{Bhattacharya:2023nmv}.
In addition to LaMET, GPDs can also be accessed on the lattice with a short-distance factorization of the spatial correlators, or the pseudo-GPD method~\cite{Radyushkin:2017cyf,Radyushkin:2019owq}, which can be used to extract the lowest few Mellin moments through operator product expansion~\cite{Izubuchi:2018srq} or fit the $x$-dependence with modeling~\cite{Bhattacharya:2024qpp}.
The ETMC in Refs.~\cite{Bhattacharya:2023ays,Bhattacharya:2024wtg} determined the moments of the unpolarized and helicity nucleon GPDs in the zero skewness case at a pion mass of 260 MeV, up to the sixth order.
The HadStruc Collaboration in Ref.~\cite{HadStruc:2024rix} also extracted the unpolarized nucleon GPD moments in this framework up to the fourth order.
Notably, according to a recent proposal by the ETMC and BNL-ANL collaboration in Ref.~\cite{Bhattacharya:2022aob}, the lattice computational cost of quasi-GPD matrix elements can be significantly reduced with the use of asymmetric frames~\cite{Bhattacharya:2023jsc,Schoenleber:2024auy}, which will improve the precision of GPD calculation in return.

The field of LaMET has matured to the point at which it is important to study and control perturbation theory uncertainties.
Much progress has been made on this front with renormalization group resummation (RGR)~\cite{Gao:2021hxl,Su:2022fiu}, leading-renormalon resummation (LRR)~\cite{Zhang:2023bxs,Holligan:2023rex} and threshold resummation~\cite{Gao:2021hxl,Ji:2023pba,Baker:2024zcd,Ji:2024hit}.
The RGR procedure is designed to resum logarithmic terms that become large when the intrinsic physical scale of the parton differs from the desired renormalization scale of the PDF.
The technique is to set the energy scale where the logarithmic terms vanish and then evolve to the desired scale using the renormalization group equation (RGE).
In the case of PDFs this is the Dokshitzer-Gribov-Lipatov-Altarelli-Parisi (DGLAP) equations~\cite{Altarelli:1977zs,Dokshitzer:1977sg,Gribov:1972ri} which have been computed up to three loops~\cite{Moch:2004pa}. 
For lightcone distribution amplitudes (DAs) it is the Efremov-Radyushkin-Brodsky-Lepage (ERBL) equation~\cite{Efremov:1978rn,Efremov:1979qk,Lepage:1979zb,Lepage:1980fj}.
The first application of RGR in the lightcone matching was to the pion valence quark PDF~\cite{Gao:2021hxl,Gao:2022iex,Su:2022fiu,Gao:2023lny,Holligan:2024umc}.
Subsequent applications of RGR matching were made to the nucleon helicity~\cite{Holligan:2024wpv} and transversity PDFs~\cite{LatticeParton:2022xsd,Gao:2023ktu}, as well as the nucleon and pion GPDs at zero skewness~\cite{Holligan:2023jqh,Ding:2024hkz}.
LRR regularizes and eliminates the infrared renormalon ambiguity between the lattice renormalization and the $\overline{\text{MS}}$ schemes, which ensures the linear power accuracy of the LaMET calculation.
It has been applied to improve the calculation of pion valence PDF~\cite{Zhang:2023bxs,Gao:2023lny,Holligan:2024umc}, light-meson DAs~\cite{Holligan:2023rex,Baker:2024zcd,Cloet:2024vbv}, nucleon transversity PDF~\cite{Gao:2023ktu}, and the pion and nucleon GPDs~\cite{Holligan:2023jqh,Ding:2024hkz}. Finally, the threshold resummation has been included in the analysis of the Mellin moments of the pion valence quark PDF~\cite{Gao:2021hxl} and pion/kaon DAs~\cite{Baker:2024zcd,Cloet:2024vbv}.

The application of RGR to DAs and GPDs at non-zero skewness, however, are more complicated, because of their multi-scale nature. In these off-forward matrix elements, there are two partons with different momenta, corresponding to two different logarithms. It is then impossible to solve just one RGE to resum their matching kernels. The soft-collinear effective theory (SCET)~\cite{Bauer:2000ew,Bauer:2001ct,Bauer:2000yr,Bauer:2001yt,Becher:2006mr} has been a useful tool to resum multi-scale problems, which further factorizes different physical scales and introduces more RGEs. It has been recently applied to quasi-PDF to resum the threshold logarithms related to soft gluon emission~\cite{Ji:2023pba,Ji:2024hit}, which becomes important when the parton momentum fraction $x\to 1$. 

In the same spirit, we seek for a further factorization of GPD matching to resum the different logarithms separately. 
In this work, we find that the large logarithms in GPD matching are important only in the threshold limit to all orders, thus could be resummed after the threshold factorization. 
Therefore, we propose to resum the GPD matching kernel in the threshold limit to improve the perturbative accuracy. 
We determine the initial scales of each RGE in the threshold factorization formula and obtain the fully resummed matching kernel. 
Finally, we numerically test our approach by applying the resummed matching kernel to a GPD model. Then, we inversely match the resulting quasi-GPD to reproduce the original GPD model. 
We demonstrate that LaMET works only for a range $x\in \mathcal{X}\equiv[-1+x_0,-\xi-x_0]\cup[-\xi+x_0,\xi-x_0]\cup[\xi+x_0,1-x_0]$, where $x_0$ is a cutoff for hard parton momentum. 
Beyond this range, the non-perturbative effect and power corrections become important. 
The fact that our final result agrees with the original GPD model also demonstrates that the potential power correction and non-perturbative effect in $x\to\pm\xi$ and $\pm1$ will not be spread out to the region $\mathcal{X}$ during the inverse matching procedure. 
This guarantees the predictive power of LaMET for $x\in\mathcal{X}$ in lattice calculations.

This paper is organized as follows.
In Sec.~\ref{Sec:Methodology}, we examine the structure of logarithms in the GPD matching kernel at non-zero skewness, and demonstrate that they are important only in the threshold limit. 
We then derive the factorization of GPD matching into the Sudakov factors and the jet function in the threshold limit using the SCET framework, which can reproduce the threshold factorizations for the quasi-PDFs~\cite{Ji:2023pba,Ji:2024hit} and quasi-DAs~\cite{Baker:2024zcd,Cloet:2024vbv}.
In Sec.~\ref{Sec:Resummation}, we derive the resummation of the Sudakov factors and the jet function in the threshold limit by solving the RGEs and determining the initial scales, and use the solutions to correct the full matching kernel. 
In Sec.~\ref{Sec:Numerics} we test our formalism numerically on a GPD model to demonstrate the self-consistency of our method.
Finally, we conclude in Sec.~\ref{Sec:Conclusion}.

\section{Threshold factorization of the quasi-GPD} 
\label{Sec:Methodology}

The process of RGR is applied to the lightcone matching. Once we have computed the quasi-GPD $\tilde{F}(x,\xi,P_z,t)$, we align the ultraviolet (UV) part with the lightcone to obtain the GPD $F(x,\xi,\mu)$ using the matching formula
\begin{equation}\label{eq:factorization}
    \tilde{F}(x,\xi,P_z,t)=\int^1_{-1}\diff{}{y}\calC\left(x,y,\xi,\frac{P_z}{\mu}\right)F(y,\xi,t,\mu)+\mathcal{O}\left(\left(\frac{\Lambda_\text{QCD}}{|x\pm\xi|P_z}\right)^2,\left(\frac{\Lambda_\text{QCD}}{|x\pm1|P_z}\right)^2\right),
\end{equation}
where $\cal C$ is the matching kernel.
For unpolarized GPDs at nonzero skewness, the matching kernel has been calculated up to NLO in Refs.~\cite{Ji:2015qla,Xiong:2015nua,Liu:2019urm,Yao:2022vtp} for the $\overline{\text{MS}}$ scheme,
\begin{align}\label{eq.NLOKernel}
    \calC&\left(x,y,\xi,\frac{P_z}{\mu}\right)=\delta(x-y)\left[1+\frac{\alpha_s(\mu)C_F}{2\pi}\left(\frac{3}{2}\ln\frac{\mu^2}{4P_z^2}+\frac{5}{2}\right)\right]\\
    &+\frac{\alpha_s(\mu)C_F}{4\pi}\left[\left(\frac{|\xi+x|}{2\xi(\xi+y)}+\frac{|\xi+x|}{(\xi+y)(y-x)}\right)\left(\ln\left(\frac{4(\xi+x)^2P_z^2}{\mu^2}\right)-1\right)\right.\nonumber\\
	&+\left.\left(\frac{|\xi-x|}{2\xi(\xi-y)}+\frac{|\xi-x|}{(\xi-y)(x-y)}\right)\left(\ln\left(\frac{4(\xi-x)^2P_z^2}{\mu^2}\right)-1\right)\right.\nonumber\\
	&\left.+\left(\left(\frac{\xi+x}{\xi+y}+\frac{\xi-x}{\xi-y}\right)\frac{1}{|x-y|}-\frac{|x-y|}{\xi^2-y^2}\right)\left(\ln\left(\frac{4(x-y)^2P_z^2}{\mu^2}\right)-1\right)\right]_+\nonumber.
\end{align}
where $\alpha_s(\mu)$ is the strong coupling at energy scale $\mu$ and $C_F$ is the quadratic Casimir for the fundamental representation of SU(3).
The plus-prescription for the second bracket, ``+'', regulates the singularity at $x=y$:
\begin{equation}
    \calC\left(x,y,\xi,\frac{P_z}{\mu}\right)_+=\calC\left(x,y,\xi,\frac{P_z}{\mu}\right)-\delta(x-y)\int\diff{}{z}\calC\left(z,y,\xi,\frac{P_z}{\mu}\right)
\end{equation}
In Eq.~\ref{eq.NLOKernel}, the quark-momentum logs are $\ln\left(\frac{4(\xi\pm x)^2P_z^2}{\mu^2}\right)$ which become large in the limit $x\to\mp \xi$ and the threshold logs are $\ln\left(\frac{4(x-y)^2P_z^2}{\mu^2}\right)$, which become large in the limit $x\to y$, the threshold limit.
The three logarithms correspond to three physical scales in the system: the outgoing and incoming quark (antiquark) momentum $|\xi+x|P_z$ and $|\xi-x|P_z$, and the gluon momentum $|x-y|P_z$. Due to its multi-scale nature, it is very difficult to naively resum the large logarithms in the traditional way by solving the RGE.

Actually, the problem can be simplified.
In the soft quark (antiquark) limit ($x\to\pm\xi$), the quark-momentum logarithms are suppressed by a factor of $|x\pm\xi|$, because to all orders of perturbation theory, these logarithms come from the loop integral involving the fermion propagator $\frac{\slashed{k}}{k^2+i\epsilon}$ attached to the Wilson line, where the fermion's longitudinal momentum is exactly the same as the parton momentum $k_z=(x\pm\xi)P_z$ by definition.
Meanwhile, the contribution from transverse Lorentz components $k_\perp$ is power-suppressed. As a result, the soft quark (antiquark) logarithm always appears as $|x\pm\xi|\ln^n |x\pm\xi|$ in the momentum-space twist-2 matching kernel, which vanishes in the soft quark (anti-quark) limit $x\to\pm\xi$, thus in general does not need to be resummed.

However, there are a few exceptions. As we will show below, in the threshold limit $x\to y$, the coefficients at all orders are fully determined by the Sudakov factor $H((x\pm\xi)P_z,\mu)$ after the threshold factorization.
The only dependence on the momentum fraction $(x\pm\xi)$ exists in the logarithm.
Thus in the threshold limit $x\to y$, the coefficient of the logarithm is finite, which could only be achieved by forming a ratio of $\frac{|x\pm\xi|}{y\pm\xi}$.
In the threshold limit $x\to y$, this ratio is $\pm1$, so the quark-momentum logarithms are still important and requires resummation.
Another exception is in the $\xi\to0$ limit, where there is a $\frac{1}{\xi}$ enhancement in the quark-momentum logarithms. If we first expand in $\xi$, the two logarithms $|\xi\pm x|\ln^n |\xi\pm x|$ of the two quark-momentum are combined, resulting in $\ln |x|$ that does not have a suppression factor when $x\to0$. It corresponds to the DGLAP logarithm in quasi-PDF. So a resummation in the threshold limit is not enough for the DGLAP region in general. Fortunately, inspired by the RGR of quasi-PDF, which resums the matching kernel by evolving from $\mu_h=2|x|P_z$~\cite{Su:2022fiu}, we find that if we first evaluate the matching kernel at $\mu_h=2|x|P_z$ for the DGLAP region $|x|>\xi$, the remaining logarithm will be of the form $\left(\frac{|x\pm\xi|}{2\xi(\xi\pm y)}+\frac{|x\pm\xi|}{\pm(\xi\pm y)(y-x))}\right)\ln^n \frac{|x\pm\xi|}{|x|}$, which becomes finite when $\xi\to 0$. After taking this step, the remaining logarithm becomes relevant only in the threshold limit.

Based on the above arguments, in the ERBL region $|x|<\xi$, the resummation of all three logarithms is only necessary in the threshold limit. In the DGLAP region $|x|>\xi$, a direct resummation in the threshold limit is not enough, but we can first choose $\mu_h=2|x|P_z$, then the resummation is only necessary in the threshold limit.

In the threshold limit, the matching kernel can be further factorized into a product of Sudakov factors $H$, and a jet function $J$~\footnote{Rigorously speaking, the jet function should be named a soft function, but we choose to follow the convention used in the literature for threshold resummations~\cite{Becher:2006mr,Becher:2007ty,Ji:2023pba}.}, as has been worked out for the PDF case~\cite{Ji:2023pba,Ji:2024hit}. Reference~\cite{Baker:2024zcd} extended the factorization to the DA case by assigning proper parton momenta to the Sudakov factors without providing a rigorous proof. In this work, we provide a SCET derivation of the general threshold factorization formula of quasi distributions, which can be reduced to the PDF, DA and GPD cases for different external states. The SCET has already been used to derive the factorization formula for a quasi transverse-momentum-dependent distribution defined from correlators fixed in the Coulomb gauge~\cite{Zhao:2023ptv}.

For a hadron moving at a large momentum $P$ along the $z$-direction, its wave function is dominated by collinear quark and gluon modes whose momentum $k^\mu$ scales as $(k^+,k^-,k_\perp)=(1,\lambda^2,\lambda)P^+$ with $\lambda\sim \Lambda_\text{QCD}/P^+ \ll 1$. When the parton momentum $xP^+$ or $xP_z$ is probed with $0<x<1$, the spectator momentum $(1-x)P^+$ is also hard, so the relevant degrees of freedom in QCD fields are collinear ($n$) and ultra soft ($us$) ones,
\begin{align}
	\psi &= \psi_n + \psi_{us} + \ldots\,,\qquad 	 A^\mu = A_n^\mu + A_{us}^\mu + \ldots\,,
\end{align}
where the quark and gluon modes scale as $\{\psi_n,\psi_{us}\}\sim \{\lambda,\lambda^3\}(P^+)^{3/2}$, $A_n^\mu= (A^+_n, A^-_n, A^\perp_n) \sim (1,\lambda^2,\lambda)P^+$, and $A^\mu_{us}\sim \lambda^2 P^+$~\cite{Bauer:2001yt,Bauer:2002uv}. To derive the collinear factorization formula for the quasi-PDF~\cite{Xiong:2013bka} that involves only one collinear momentum scale $xP^+$, there is no need to separate the collinear and ultrasoft degrees of freedom through field redefinition~\cite{Bauer:2001yt}.

The quark bilinear operator used to define the quasi-distributions,
\begin{align}
    O_{\Gamma}(z_1,z_2) &\equiv \bar{\psi}(z_1) \Gamma W(z_1,z_2) \psi(z_2)\,,
\end{align}
can be re-expressed as the product of two dressed quark fields~\cite{Dorn:1986dt,Ji:2017oey,Green:2017xeu},
\begin{align}
    O_{\Gamma}(z_1,z_2) &= \bar{\psi}(z_1) W_{z}(z_1) \Gamma W_{z}^\dagger (z_2) \psi(z_2) \equiv \bar{\Psi}(z_1) \Gamma \Psi(z_2)\,,
\end{align}
where $\Gamma=\gamma^t$ or $\gamma^z$, and $W_z(z)$ is an semi-infinite Wilson line that connects $\pm\infty$ to $z^\mu$. For collinear quasi-distributions, the product of $W_{z}(z_1)$ and $W_{z}^\dagger (z_2)$ reduces to the straight Wilson line from $z_2$ to $z_1$ regardless of their orientations, which is why we do not specify the direction of $W_z$ in the above equations. In the following we choose $\Gamma=\gamma^t$ for discussion.

Since $\bar{\psi}_n\gamma^- \psi_n \sim O(\lambda^4)$, $\bar{\psi}_n\gamma^+ \psi_n \sim O(\lambda^2)$, the expansion of $O_{\gamma^t}(z_1,z_2)$ is
\begin{align}\label{eq:colexp}
	O_{\gamma^t}(z_1,z_2) = {1\over \sqrt{2}}\bar{\psi}_n(z_1) W_z(z_1) \gamma^+ W_z^\dagger(z_2) \psi_n(z_2) + O(\lambda^4)\,.
\end{align}
When away from the threshold region, i.e., the emission by the active parton is also collinear and hard, the Wilson line $W_z$ can be expanded as
\begin{align}\label{eq:wexp}
    W_z = W_z[A_n] = W_n[A_n] + O(\lambda^2)\,.
\end{align}
As a result, the dressed quark field becomes
\begin{align}
    W_z^\dagger\psi_n = W_n^\dagger \psi_n + O(\lambda^2)\,.
\end{align}
In SCET, the r.h.s. corresponds to the operator
\begin{align}
        e^{-i{\cal P}\cdot z} W^\dagger_n \xi_n \,,
\end{align}
where ${\cal P}^\mu$ is the Hermitian derivative operator that projects out the collinear momentum of $W^\dagger_n \xi_n$, and $\xi_n$ is the collinear quark field with the zero mode subtracted~\cite{Manohar:2006nz}. The QCD operator $W_z^\dagger\psi_n$ and SCET operator $e^{-i{\cal P}\cdot z} W^\dagger_n \xi_n$ are related by a matching condition
\begin{align}\label{eq:scet_match}
	W_z^\dagger \psi_n(z) &= e^{-i{\cal P}\cdot z}H({\cal P}_z,\mu)W_n^\dagger\xi_n\,,
\end{align}
where $H({\cal P}_z,\mu)$ is a hard Sudakov factor independent of the quark flavor.

Since $z$ is the Fourier conjugate to $xP_z$, it scales as $z\sim O(1)$ at finite $x$. This means that the exchange of particles between the dressed fields $\bar{\psi}_n(z)W_z(z)$ and $W_z^\dagger(0)\psi_n(0)$ also contribute to the hard coefficient function, so the matching of $O_{\gamma^t}(z_1,z_2)$ from QCD to SCET must be in a convolutional form,
\begin{align}\label{eq:fact0}
	O_{\gamma^t}(z_1,z_2) &=  \int d\eta_1 d\eta_2  \Big[\bar{\xi}_nW_n \Big](z_1)  {\gamma^+  \over \sqrt{2}} \mathbf{C}(\eta_1,\eta_2, {\cal P}_1^{+\dagger}/\eta_1, {\cal P}_2^+/\eta_2, \mu)\Big[ W_n^\dagger\xi_n\Big](z_2) \nonumber\\
    &= \int d\eta_1 d\eta_2 \int d\omega_1 d\omega_2 \ \mathbf{C}(\eta_1,\eta_2, \omega_1,\omega_2,\mu) \nonumber\\
    &\qquad\times {e^{i({\cal P}_1\cdot z_1-{\cal P}_2\cdot z_2)}\over \sqrt{2}} \Big[\bar{\xi}_nW_n \Big]\delta({\omega_1}-{\cal P}^{+\dagger}_1/\eta_1)   \gamma^+ \delta({\omega_2}- {\cal P}^+_2/\eta_2)\Big[ W_n^\dagger\xi_n\Big]\,,
\end{align}
where $\mathbf{C}$ is a perturbative matching coefficient that depends on the momenta $(\omega_1,\omega_2)$ carried by each parton. The convolution is realized through the integration over the momentum fractions of the daughter partons, $\eta_1,\eta_2\in (-\infty,\infty)$~\cite{Xiong:2013bka}, as the convolution allows all collinear momenta to flow out of the active parton. Note that here we have neglected the mixing with gluon bilinears since we only consider non-singlet and valence quark channels, but it will be straightforward to include them. The variables $\eta_1,\eta_2$ will eventually be fixed by momentum conservation in the matrix elements, as we show below. The Sudakov factors $H$ are absorbed into the matching coefficient $\mathbf{C}$ and cannot be separated from the other contributions under collinear factorization.


When inserted into the off-forward hadron states, we can derive the factorization formula for the quasi-GPD,
\begin{align}
    \tilde F(x,\xi, P_z, t) &\equiv \int {dz\over 4\pi} e^{-ix P_z z}\left\langle p''\left| O_{\gamma^t}\left(-{z\over 2},{z\over2}\right) \right| p'\right\rangle \nonumber\\
    &= \int d\eta_1 d\eta_2 \int d\omega_1 d\omega_2 \ \mathbf{C}(\eta_1,\eta_2, \omega_1,\omega_2,\mu) \nonumber\\
    &\quad\times \left\langle p''\left|\Big[\bar{\xi}_nW_n \Big]\delta({\omega_1}-{\cal P}^{+\dagger}_1/\eta_1) {\delta(xP_z-{{\cal P}^{z\dagger}_1+{\cal P}^z_2\over 2})\over \sqrt{2}}   {\gamma^+\over2} \delta({\omega_2}- {\cal P}^+_2/\eta_2)\Big[ W_n^\dagger\xi_n\Big]\right|p'\right\rangle \nonumber\\
    &= \int d\eta_1 d\eta_2 \int d\omega_1 d\omega_2 \ \mathbf{C}(\eta_1,\eta_2, \omega_1,\omega_2,\mu) \nonumber\\
    &\quad\times \left\langle p''\left|\Big[\bar{\xi}_nW_n \Big]\delta({\omega_1}-{\cal P}^{+\dagger}_1/\eta_1) \delta(xP^+-{{\cal P}^{+\dagger}_1+{\cal P}^+_2\over 2})  {\gamma^+\over2} \delta({\omega_2}- {\cal P}^+_2/\eta_2)\Big[ W_n^\dagger\xi_n\Big]\right|p'\right\rangle \nonumber\\
    &=  \int d\eta_1 d\eta_2 \int d\omega_1 d\omega_2 \ \mathbf{C}(\eta_1,\eta_2, \omega_1,\omega_2,\mu) \delta\left(xP^+-{1\over 2}{\left({\omega_1 \eta_1} + {\omega_2 \eta_2}\right)}\right) \nonumber\\
    &\quad \times   P^+ \int_{-1}^1 dy\ \delta(\omega_1- (y+\xi) P^+) \delta(\omega_2- (y-\xi)P^+) F(y,\xi,t,\mu)\nonumber\\
    &=\int_{-1}^1 dy \int d\eta_1 d\eta_2 \ \mathbf{C}(\eta_1,\eta_2, (y+\xi) P^+,(y-\xi) P^+,\mu) \nonumber\\
    &\quad \times \delta\left(x-{1\over 2}{\left((y+\xi)\eta_1 + (y-\xi)\eta_2\right)}\right)  F(y,\xi,t,\mu)\,.
\end{align}

If we perform a change of variables,
\begin{align}
    \bar{\eta} &= {1\over 2} \left[(1+\rho) \eta_1 +(1-\rho) \eta_2 \right]\,,\qquad  \eta =-{1-\rho \over 1+\rho^2} \eta_1 + {1+\rho \over 1+\rho^2}\eta_2\,,
\end{align}
with $\rho = {\xi\over y}$, then
\begin{align}
    \tilde F(x,\xi, P_z, t) &= \int_{-1}^1 {dy\over |y|} \int d\eta \ \mathbf{C}\left(\bar{\eta}={x\over y},\eta,\rho,(y+\xi) P^+,(y-\xi) P^+,\mu\right) F(y,\xi,t,\mu)\,.
\end{align}
As one can see, $\eta$ is merely a dummy variable to be integrated over, and the matching coefficient is uniquely determined by the momenta of the active parton and its emission. Therefore, by defining
\begin{align}
   {1\over |y|} \int d\eta \ \mathbf{C}\left({x\over y},\eta,{\xi\over y}, (y+\xi) P^+,(y-\xi) P^+,\mu\right) & \equiv {\cal C}\left({x\over y},{\xi\over y},(y+\xi) P^+,(y-\xi) P^+,\mu\right)_+\nonumber\\
    &= \calC\left(x,y,\xi,\frac{P_z}{\mu}\right)_+\,,
\end{align}
we obtain the exact same factorization formula Eq.~\eqref{eq:factorization} derived in Ref.~\cite{Liu:2019urm} using operator product expansion. 
Based on Eq.~\eqref{eq:fact0}, we can also derive the collinear factorization formulae for the quasi-PDF~\cite{Xiong:2013bka,Ma:2014jla,Izubuchi:2018srq} and quasi-DA~\cite{Liu:2018tox}.

Now let us turn to the threshold limit $x\to y$, where we consider a ``soft'' scale $|x-y|P^+ \sim \sqrt{\epsilon} P^+ \ll P^+$ for the emission from the active parton. This scale is soft ($\epsilon \ll 1$) compared to the collinear momentum, while it can still be much harder or comparable to $\lambda P^+$, i.e., $\Lambda_\text{QCD}$. When $\sqrt{\epsilon} P^+\sim \Lambda_\text{QCD}$, then perturbation theory fails and the factorization will involve a non-perturbative jet function. Therefore, we consider $\sqrt{\epsilon} P^+\gg \Lambda_\text{QCD}$ so that the jet function is still perturbatively calculable~\cite{Ji:2023pba,Ji:2024hit}. After threshold resummation, we shall be able to see when perturbation theory breaks down.


Under this hierarchy, the emitted ``soft''-collinear modes $\sqrt{\epsilon}(1,\lambda^2,\lambda)P^+$ from the active parton can propagate a distance of $1/(\sqrt{\epsilon} P^+)$, which makes the ``soft'' degrees of freedom $(\sqrt{\epsilon},\sqrt{\epsilon},\sqrt{\epsilon})P^+$ relevant. To derive the threshold factorization formula, we first integrate out the off-shell modes with $p^2\gg\epsilon (P^+)^2$ to match QCD to a SCET which involves the ``collinear'' $\sim(1,\epsilon,\sqrt{\epsilon})P^+$ and ``soft'' modes $\sim(\sqrt{\epsilon},\sqrt{\epsilon},\sqrt{\epsilon})P^+$. The corresponding ``soft-collinear'' mode, which scales as $\sim \sqrt{\epsilon}(1, \epsilon, \sqrt{\epsilon})P^+$ and is emitted by the parent parton, is also relevant. If treated separately from the ``collinear'' mode, it would lead to a further factorization of the PDF into a ``hard-collinear'' coefficient and a ``soft'' function~\cite{Korchemsky:1992xv,Becher:2006mr,Ji:2024hit}. However, since our goal is to obtain the full PDF without additional factorization, we do not perform this separation, and the ``soft-collinear'' contribution is included as part of the ``collinear'' modes.
With such a scale separation, the space-like Wilson line in the dressed quark field follows the factorized expansion,
\begin{align}
    W_z &= W_z[A_n+A_s] = W_n[A_n] S_z[A_s] + O(\epsilon)\,,
\end{align}
after integrating out the off-shell modes with $p^2 \gg \sqrt{\epsilon}(P^+)^2$ that couple the ``collinear'' and ``soft'' modes. The ``soft'' Wilson line $S_z[A_s]$ cannot be expanded in the collinear direction because the gluon momentum is off-shell and isotropic. On the other hand, the contribution from ``soft'' quark modes in the bilinear operator $O_{\gamma_t}$ is suppressed and negligible. Therefore, we have the collinear expansion
\begin{align}
    O_{\gamma^t}(z_1,z_2) &={1\over \sqrt{2}} \bar{\psi}_n(z_1) W_n(z_1) S_z(z_1) \gamma^+ S^\dagger_z(z_2)  W_n(z_2) \psi_n(z_2) + O(\epsilon^2)\,.
\end{align}

Since we consider the distance $|z_1-z_2|\sim O(1/(\sqrt{\epsilon}P^+)$ propagated by the ``soft'' modes, the ``collinear'' fields separated by it will not be able to exchange any hard particle. Therefore, the QCD operator is matched onto the above SCET as
\begin{align}\label{eq:fact1}
    O_{\gamma^t}(z_1,z_2) &={ e^{i{\cal P}_1\cdot z_1 - {\cal P}_2\cdot z_2}\over \sqrt{2}} \left[\bar{\xi}_nW_n\right]_\epsilon H({\cal P}_1^{z\dagger},\mu)S_n^\dagger(z_1) S_z(z_1)\nonumber\\
    &\qquad \times \gamma^+S_z^\dagger(z_2)S_n(z_2) H({\cal P}_2^z,\mu) \left[W_n^\dagger \xi_n \right]_\epsilon + O(\epsilon^2)\,,
\end{align}
where the matching for the collinear fields is multiplicative and given by the Sudakov factors. The soft Wilson lines $S_n$ is added to ensure invariance under the collinear and soft gauge transformations in SCET~\cite{Bauer:2001yt}. We also label the square brackets enclosing the collinear fields by ``$\epsilon$'' to distinguish them from the original SCET defined by the expansion parameter $\lambda$.

The hadron matrix element of the ``collinear'' fields defines the GPD, which is equivalent to the definition in full QCD under lightcone quantization or the original SCET~\cite{Bauer:2001yt,Becher:2007ty}.
Therefore, we can remove the label $\epsilon$ and establish the threshold factorization formula for the quasi-GPD as
\begin{align}\label{eq:fact2}
    \tilde F(x,\xi, P_z, t) &= \int {dz\over 4\pi} e^{-ix P_z z}\left \langle p''\left| \left[\bar{\xi}_nW_n\right] H({\cal P}_1^{z\dagger},\mu)S_n^\dagger(z_1) S_z(z_1) {e^{i{{\cal P}^{z\dagger}_1+{\cal P}^z_2\over 2} \cdot z}\over \sqrt{2}}\right.\right. \nonumber\\
    &\qquad \left.\left.\times {\gamma^+\over 2} S_z^\dagger(z_2)S_n(z_2) H({\cal P}_2^z,\mu) \left[W_n^\dagger \xi_n \right] \right| p \right\rangle \nonumber\\
    &=\int {dz\over 4\pi} e^{-ix P_z z} \left\langle p''\left| \left[\bar{\xi}_nW_n\right] H({\cal P}_1^{z\dagger},\mu) {e^{i{{\cal P}^{z\dagger}_1+{\cal P}^z_2\over 2} \cdot z}\over \sqrt{2}}  {\gamma^+\over 2}  H({\cal P}_2^z,\mu) \left[W_n^\dagger \xi_n \right] \right|p'\right\rangle\nonumber\\
    &\qquad \times {1\over N_c} \left\langle 0\left| \Tr\left[S_n^\dagger(-{z\over2}) S_z(-{z\over2}) S_z^\dagger({z\over2})S_n({z\over2})\right] \right|0\right\rangle \,,
\end{align}
where we derive the coordinate-space jet function definition
\begin{align}
    \tilde{J}_z(z^2\mu^2) &\equiv {1\over N_c} \left\langle 0\left| \Tr\left[S_n^\dagger(-{z\over2}) S_z(-{z\over2}) S_z^\dagger({z\over2})S_n({z\over2})\right] \right|0\right\rangle\,.
\end{align}

The collinear matrix element in the above factorization formula defines the lightcone GPD,
\begin{align}\label{eq:fact3}
    \left\langle p''\left| \left[\bar{\xi}_nW_n\right] H({\cal P}_1^{z\dagger},\mu) {\gamma^+\over 2}  H({\cal P}_2^z,\mu) \left[W_n^\dagger \xi_n \right] \right|p'\right\rangle &= \int_{-1}^1 dy \ H((y-\xi)P_z,\mu)^\dagger  H((y+\xi)P_z,\mu) \nonumber\\
    &\qquad\times F(y,\xi,t,\mu) \,.
\end{align}
However, additional care must be given to the Sudakov factors. When one matches the dressed QCD field $W_z^\dagger \psi$ to the SCET operator $W_n^\dagger \xi_n$, the Sudakov factor also involves an imaginary part~\cite{Ji:2019ewn,Ji:2021znw,Vladimirov:2020ofp,Avkhadiev:2023poz},
\begin{align} \label{eq:sudakov}
    H({\cal P}^z/\mu) &= H\left(\ln{-4({\cal P}^z\pm i0)^2\over \mu^2}\right)\,,
\end{align}
where $\pm$ depends on the orientation of the Wilson lines.

As we have explained in the above, the quasi-GPD does not depend on the orientation of the Wilson line $W_z$, so the threshold factorization should not depend on it, either. However, this apparently contradicts Eqs.~\eqref{eq:fact3} and \eqref{eq:sudakov}, as there is an incomplete cancellation of the imaginary parts of the two Sudakov factors, which leads to a remnant contribution that depends on the orientation of $W_z$. To resolve this conflict, we note that the key in the threshold factorization is that the collinear fields separated by $|z|\sim 1/(\sqrt{\epsilon}P^+)$ do not exchange hard particles. Therefore, in the spacetime picture of the factorization, the dressed quark fields $\bar{\psi}W_z$ and $W_z^\dagger \psi$ must already be spatially separated, which cannot be satisfied if both $W_z$'s have the same orientation. In Refs.~\cite{Ji:2023pba,Ji:2024hit}, it was proposed to choose $W_z$ differently for $\bar{\psi}$ and $\psi$, i.e., $\big[\bar{\psi}W_{-\hat{z}}\big](-z/2)$ and $\big[W_{\hat{z}}^\dagger \psi\big](z/2)$ for $z>0$, and $\big[\bar{\psi}W_{\hat{z}}\big](-z/2)$ and $\big[W_{-\hat{z}}^\dagger \psi\big](z/2)$ for $z<0$. In this way, the Wilson lines avoid overlapping with each other, and the definition of the jet function should be modified accordingly to involve both future- and past-pointing light-like Wilson lines. As a result, the imaginary part of the Sudakov factor depends on the sign of $z$, which can be redefined with a phase,
\begin{align}\label{eq:phase}
    H({\cal P}^{z},\mu,\pm) &=H\left(\ln{-4({\cal P}^z\pm i\sgn(z)0)^2\over \mu^2}\right)\equiv |H({\cal P}^z,\mu)| \exp\left[\mp i{\sgn(z{\cal P}^z)}A({\cal P}^z/\mu)\right]\,.
\end{align}
Nevertheless, such a treatment has a problem of recovering the straight Wilson line in the quasi-GPD definition. To reconcile this inconsistency, we note that when both $W_z$'s are of the same orientation, they overlap with each other, so there can be exchange of hard particles between the dressed quarks. Although they are separated by $|z|\sim 1/(\sqrt{\epsilon}P^+)$, there is still a segment of the straight Wilson line $W(-z/2,z/2)$ that is within a distance $\sqrt{\epsilon}|z|$ of $\bar{\psi}$ or $\psi$, over which the hard particles can travel. The contribution from this segment is exactly the so called ``sail diagram''~\cite{Izubuchi:2018srq} in the threshold limit $x\to y$, which is identical to the product of the phase factors in Eq.~\eqref{eq:phase} and independent of the choice of $W_z$ orientation.

Therefore, the exact factorization formula for the quasi-GPD is established as
\begin{align}    
    \tilde F(x,\xi, P_z, t) &= \int_{-1}^1 dy F(y,\xi,t,\mu)\! \int {dz\over 4\pi} e^{-i(x-y)P_z z}H((x\!-\!\xi)P_z,\!\mu,\!-\!)^\dagger H((x\!+\!\xi)P_z,\!\mu,\!+\!)  \tilde J_z(z^2\mu^2) \nonumber\\
    &=\int_{-1}^1 dy|H((x-\xi)P_z,\mu) H((x+\xi)P_z,\mu)| F(y,\xi,t,\mu)\int {dz\over 4\pi} e^{-i(x-y)P_z z} \tilde{J}_z(z^2\mu^2) \nonumber\\
    &\qquad \times \exp\left[-i{\sgn(z(x-\xi)P_z)}A((x-\xi)P_z/\mu)-(\xi\to -\xi)\right]\,.
\end{align}
Note that in the first line we have replaced the variable $y$ by $x$ in the Sudakov factors, which is allowed since we work in the limit $x\to y$. The above factorization formula reduces to the PDF case~\cite{Ji:2023pba,Ji:2024hit} in the limit $\xi\to0$. For the DA case, the leading Fock state of the meson is a pair of quark and antiquark with momentum fraction $x$ and $(1-x)$, respectively. Therefore, we can define the antiquark momentum to be negative in the Sudakov factor, i.e., $H(-(1-x)P^z,\mu,-)^\dagger H(xP_z,\mu,+) $, to obtain the threshold factorization formula~\cite{Baker:2024zcd,Cloet:2024vbv}. 

Therefore, the matching coefficient in the threshold limit is factorized as
\begin{align}\label{eq.TRFactorization}
    \calC\left(x,y,\xi,\frac{P_z}{\mu}\right)& \xrightarrow{x\to y}{\cal F} \left[H((x-\xi)P_z,\mu,-)^\dagger H((x+\xi)P_z,\mu,+)\right]\nonumber\\
    &\qquad\qquad \otimes J(|x-y|P_z,\mu)(1+\mathcal{O}(x-y)),
\end{align}
where ${\cal F}$ stands for Fourier transform from $z$ to $x-y$.

The Sudakov factor has the following form in coordinate space up to one-loop order,
\begin{align}
    H(p_z,\mu,\pm)&=1+\frac{\alpha_s(\mu)C_F}{4\pi}\left(-\frac{1}{2}\mathbf{L}_{\pm}^2(p_z)+\mathbf{L}_{\pm}(p_z)-\frac{5\pi^2}{12}-2\right),\\
    \mathbf{L}_{\pm}(p_z)&=\ln\left(\frac{-4(p_z\pm i\sgn(z)0)^2}{\mu^2}\right)=\ln\frac{4p_z^2}{\mu^2}\mp i\sgn(zp_z)\pi.
\end{align}
After Fourier transformation, the real part is independent of $z$, thus goes into a delta function, and the imaginary part's $z$-dependence in sign$(zp_z)$ is converted to $\frac{1}{(y-x)\pi}$,
\begin{align}
    \mathcal{F}[\Re(H(p_z,\mu,\pm))](x-y)&=\delta(x-y)\left[1+\frac{\alpha_s(\mu)C_F}{4\pi}\left(-\frac{1}{2}\ln^2\frac{4p_z^2}{\mu^2}+\ln\frac{4p_z^2}{\mu^2}+\frac{\pi^2}{12}-2\right)\right],\nonumber\\
    \mathcal{F}[\Im(H(p_z,\mu,\pm))](x-y)&=\mp \sgn(zp_z)\frac{\alpha_s(\mu)C_F}{4\pi}\frac{1}{y-x}\left(\ln\frac{4p_z^2}{\mu^2}-1\right).
\end{align}

The jet function is derived from the Wilson line structure and is thus the same for GPDs as it is for PDFs. In coordinate space, the jet function is
\begin{align}
    \tilde{J}_z(z,\mu)=1+\frac{\alpha_s(\mu)C_F}{2\pi}\left(\frac{1}{2}l_z^2+l_z+\frac{\pi^2}{12}+2\right),
\end{align}
where $l_z=\ln \left(\frac{\mu^2z^2e^{2\gamma_E}}{4}\right)$. The corresponding momentum-space formalism is
\begin{align}
    J(|x-y|P_z,\mu)& =\delta(x-y)\left(1+\frac{\alpha_s(\mu)C_F}{2\pi}\left(2+\frac{\pi^2}{4}+\frac{1}{2}\ln^2\frac{4P_z^2}{\mu^2}-\ln\frac{4P_z^2}{\mu^2}\right)\right)\nonumber\\
    &+\frac{\alpha_s(\mu)C_F}{2\pi}\left(\mathcal{P}\left(\frac{\ln(x-y)^2}{|x-y|}\right)+\left(\ln\frac{4P_z^2}{\mu^2}-1\right)\mathcal{P}\left(\frac{1}{|x-y|}\right)\right),
\end{align}
with the principal value (PV) $\mathcal{P}(f(|x-y|))$ defined as
\begin{align}
    \mathcal{P}(f(|x-y|))=f(|x-y|)-\delta(x-y)\int_{x-1}^{x+1} dy f(|x-y|).
\end{align}

In practice, lattice quasi-GPDs are renormalized in the hybrid scheme~\cite{Ji:2020brr}, which is perturbatively convertible to $\overline{\text{MS}}$. For example, The multiplicative renormalization factor in coordinate space can be defined as
\begin{align}
Z^\text{hybrid}(z,a)=\left\{
\begin{matrix}
\langle P=0| O_\Gamma(0,z)|P=0\rangle,\hfill  &  |z|\leq z_s \\
\langle P=0| O_\Gamma(0,z_s)|P=0\rangle e^{-\delta m(a) (|z|-z_s)}, & |z|>z_s
\end{matrix}
\right.
\label{eq:hybrid_scheme}
\end{align}
where $\langle P=0| O_\Gamma(0,z)|P=0\rangle$ is the matrix element of operator $O_\Gamma$ in $P=0$ hadron states, and $\delta m(a)$ is the mass counterterm to remove the linear divergence of the spatial Wilson line in $O_\Gamma$~\cite{Chen:2016fxx,Green:2017xeu,Ji:2017oey,Ishikawa:2017faj,Ji:2020brr}.
The matching kernel then requires a modification to the $\overline{\text{MS}}$ matching kernel Eq.~\eqref{eq.NLOKernel} as~\cite{Ji:2020brr}
\begin{align}\label{eq:hybrid_matching}
    \Delta\mathcal{C}(x,y,\xi,\frac{P_z}{\mu})=\frac{\alpha_s(\mu)C_F}{2\pi}\left[-\delta(x-y)\left(\frac{3}{2}\ln\frac{\mu^2}{4P_z^2}+\frac{5}{2}\right)+\left(\frac{3\Si[(y-x)z_sP_z]}{\pi(y-x)}\right)_+\right],
\end{align}
where $z_s\ll \Lambda_\text{QCD}^{-1}$ separates short and long range renormalizations, and $\Si(\lambda)=\int_0^\lambda dt\sin(t)/t $. Since the singular terms in the threshold limit has been changed by this correction term, the threshold factorization will also be modified. As suggested in Ref.~\cite{Ji:2024hit}, it is convenient to absorb the correction into the norm of the Sudakov factor,
\begin{align}\label{eq:hybrid_corr}
    \Delta |H(p_z,\mu)|&=-\frac{\alpha_s(\mu)C_F}{4\pi}\left(\frac{3}{2}\ln\frac{\mu^2z_s^2e^{2\gamma_E}}{4}+\frac{5}{2}\right),
\end{align}
which reproduces the threshold terms of $\Delta\mathcal{C}$ in momentum space.

\section{Resumming large logarithms in the threshold limit} 
\label{Sec:Resummation}
\subsection{Renormalization group equation and the general solutions}
The lightcone GPD follows an evolution equation
\begin{align}\label{eq:gpd_evo}
    \frac{\partial F(x,\xi,t)}{\partial \ln\mu^2}&=\hat{\mathcal{V}}F= \mathcal{V}(x,y,\xi)\otimes F(y,\xi,t)\,,\\
    \mathcal{V}(x,y,\xi) &=\frac{\alpha_s(\mu)C_F}{4\pi}\left[\left(\frac{|\xi+x|}{2\xi(\xi+y)}+\frac{|\xi+x|}{(\xi+y)(y-x)}\right)+\left(\frac{|\xi-x|}{2\xi(\xi-y)}+\frac{|\xi-x|}{(\xi-y)(x-y)}\right)\right.\nonumber\\
	&\qquad\left.+\left(\left(\frac{\xi+x}{\xi+y}+\frac{\xi-x}{\xi-y}\right)\frac{1}{|x-y|}-\frac{|x-y|}{\xi^2-y^2}\right)\right]_++\mathcal{O}(\alpha_s^2),
\end{align}
which allows us to evolve the GPD from one fixed scale to another. It is not enough to resum all three different logarithms with this single RG evolution equation. But as we mentioned in the previous section, the threshold factorization split the three different logarithms into three different quantities.
Now all three factorized parts of the matching kernel follow independent RGEs. This allows the separate resummation of the three components. More explicitly, the
 Sudakov factors follow the RG equation
\begin{align}\label{eq:sudakov_evolution}
    \frac{\partial\ln H(p_z,\mu,\pm)}{\partial\ln\mu}=\frac{1}{2}\Gamma_\text{cusp}(\alpha_s)\left(\ln\frac{4p_z^2}{\mu^2}\mp i \pi \sgn(z p_z)\right)+\gamma_H(\alpha_s),
\end{align}
where $\Gamma_\text{cusp}$ is the universal cusp anomalous dimension~\cite{Korchemsky:1987wg} known to four-loop order~\cite{Henn:2019swt,vonManteuffel:2020vjv},
\begin{align}
    \Gamma_\text{cusp}&=\frac{4\alpha_s}{3\pi}+\frac{\alpha_s^2}{27\pi^2}\left[201-9\pi^2-10n_f\right]+\frac{\alpha_s^3}{3240\pi^3}\left[99225 + 330 n_f - 40 n_f^2 - 12060 \pi^2 \right.\nonumber\\
    &\left. + 600 n_f \pi^2 +  594 \pi^4 + 17820 \zeta(3)- 13320 n_f \zeta(3)\right]+\mathcal{O}(\alpha_s^4),\nonumber
\end{align}
where $\zeta(3)$ is the Riemann zeta function,  $n_f$ is the number of active quark flavors, and $\gamma_H$ is the anomalous dimension of the Sudakov factor known to 2-loop order~\cite{Ji:2021znw,Ji:2023pba,delRio:2023pse}:
\begin{align}
    \gamma_H=&-\frac{2\alpha_s}{3\pi}+\frac{\alpha_s^2}{648\pi^2}\left[1836\zeta(3)+39\pi^2-3612+(160+18\pi^2)n_f\right]+\mathcal{O}(\alpha_s^3). \nonumber
\end{align}
In the hybrid scheme, the anomalous dimension of the Sudakov factor is modified correspondingly,
\begin{align}
    \Delta \gamma_H=-\frac{\alpha_s}{\pi}-{\alpha_s^2}\left(\frac{127}{72\pi^2}+\frac{7}{54}-\frac{5n_f}{36\pi^2}\right).
\end{align}
The norm and phase of the Sudakov factor follow independent evolutions,
\begin{align}
\frac{\partial\ln |H|(p_z,\mu)}{\partial\ln\mu}&=\frac{1}{2}\Gamma_\text{cusp}(\alpha_s)\ln\frac{4p_z^2}{\mu^2}+\gamma_H(\alpha_s),\\
    \frac{\partial A(p_z,\mu)}{\partial\ln\mu}&=\pi\Gamma_\text{cusp}.
\end{align}
Both of them can be solved analytically as
\begin{align}
|H|^\text{TR}(p_z,\mu)&=|H|(\mu_{h})e^{S_\Gamma(\mu_h,\mu)-a_H(\mu_h,\mu)}\left(\frac{2p_z }{\mu_h}\right)^{-a_\Gamma(\mu_h,\mu)},\\
A(p_z,\mu)&=A(p_z,\mu_h)+\pi a_\Gamma(\mu_h,\mu),
\end{align}
where $S_\Gamma(\mu_0,\mu)$, $a_H(\mu_0,\mu)$ and $a_\Gamma(\mu_0,\mu)$ are all evolution factors from scale $\mu_0$ to $\mu$, calculated from the QCD beta function and the anomalous dimensions, 
\begin{align}
    &S_\Gamma(\mu_0,\mu)=-\int_{\alpha_s(\mu_0)}^{\alpha_s(\mu)}\frac{\Gamma_\text{cusp}(\alpha)d\alpha}{\beta(\alpha)}\int_{\alpha_s(\mu_0)}^{\alpha}\frac{d\alpha'}{\beta(\alpha')},\nonumber\\
    &a_H(\mu_0,\mu)=-\int_{\alpha_s(\mu_0)}^{\alpha_s(\mu)}\frac{\gamma_{H}(\alpha)d\alpha}{\beta(\alpha)},\\
    &a_{\Gamma}(\mu_0,\mu)=-\int_{\alpha_s(\mu_0)}^{\alpha_s(\mu)}\frac{\Gamma_\text{cusp}(\alpha)d\alpha}{\beta(\alpha)}.\nonumber
\end{align}

The jet function evolves multiplicatively in coordinate space,
\begin{align}\label{eq:jet_evolution}
    \frac{\partial\ln \tilde{J}(z,\mu)}{\partial\ln\mu}=\Gamma_\text{cusp}(\alpha_s)l_z-\gamma_J(\alpha_s),
\end{align}
where $\gamma_J$ is known at up to two-loop order~\cite{Ji:2023pba},
\begin{align}
    \gamma_J=&-\frac{4\alpha_s}{3\pi}+\frac{\alpha_s^2}{12\pi^2} \left[60\zeta(3)+\frac{23\pi^2}{3}-\frac{1396}{9}+(\frac{233}{27}-\frac{2\pi^2}{9})n_f\right]+\mathcal{O}(\alpha_s^3).\nonumber
\end{align}
It has a simple solution,
\begin{align}
\tilde{J}_z(l_z,\mu)=e^{\left[-2S_\Gamma(\mu_i,\mu)+a_J(\mu_i,\mu)\right]}\left(\frac{z\mu_i e^{\gamma_E}}{2}\right)^{-2a_\Gamma(\mu_i,\mu)}\tilde{J}_z(l_z,\alpha_s(\mu_i)),
\end{align}
with the evolution factor
\begin{align}
    &a_J(\mu_i,\mu)=-\int_{\alpha_s(\mu_i)}^{\alpha_s(\mu)}\frac{\gamma_{J}(\alpha)d\alpha}{\beta(\alpha)}.
\end{align}
Its Fourier conjugate
$J(\Delta=|x-y|P_z,\mu)$ evolves in a more complicated way,
\begin{align}
    \frac{\partial J(\Delta,\mu)}{\partial\ln\mu}=&-\left[\Gamma_\text{cusp}(\alpha_s)+\gamma_J(\alpha_s)\right]J(\Delta,\mu)
    \nonumber\\
    &+\Gamma_\text{cusp}(\alpha_s)\left(\int_{ \Delta {}'< \Delta } d\Delta{}' \frac{J(\Delta{}',\mu)}{\Delta-\Delta{}'}+\int_{\Delta{}'>\Delta} d\Delta{}' \frac{J(\Delta{}',\mu)}{\Delta+\Delta{}'}\right),
\end{align}
so it is more straightforward to resum the jet function in coordinate space, then Fourier transform back to momentum space~\cite{Becher:2006mr},
\begin{align}
\label{eq:jet_fun_resum}
J(\Delta,\mu)=e^{\left[-2S_\Gamma(\mu_i,\mu)+a_J(\mu_i,\mu)\right]}\tilde{J}_z(l_z=-2\partial_\eta,\alpha_s(\mu_i))&\left.\left[\frac{\sin(\eta\pi/2)}{|\Delta|}\left(\frac{2|\Delta|}{\mu_i}\right)^{\eta}\right]_*\right.\nonumber \\
&
\times\left.\frac{\Gamma(1-\eta)e^{-\eta\gamma_E}}{\pi}\right|_{\eta=2a_{\Gamma}(\mu_i,\mu)},
\end{align}
where the star function is a generalized version of the plus function,
\begin{align}
    \int_{-\infty}^{\infty} d\Delta&\left[|\Delta|^{\eta-1}\right]_* f(\Delta) \equiv\int_{-\infty}^{\infty} d\Delta |\Delta|^{\eta-1}\left(f(\Delta)-\sum_{i=0}^{\lfloor -\eta\rfloor} \frac{\Delta^i}{i!} f^{(i)}(0)\right),
\end{align}
with $\lfloor -\eta\rfloor=n$ when $n\leq -\eta < n+1$ for an integer $n$.

\subsection{Initial-scale choice for the solutions to the RGEs}

Although we have the general solutions to the RGEs, we still need to figure out the initial scales, $\mu_{h_1}$, $\mu_{h_2}$ and $\mu_i$ to obtain the resummed forms.
Despite the explicit dependence on these initial scales in the formalism, the solutions to the RGEs are actually independent of them when summed up to all orders of $\alpha_s$,
\begin{align}
    \frac{\partial J(\Delta,\mu)}{\partial \mu_i}=\frac{\partial H(p_z,\mu)}{\partial \mu_h}=0.
\end{align}
However, since we truncate the perturbation series up to a fixed-order of $\alpha_s$, and we only aim to resum the logarithms at higher order, there are still remaining initial-scale dependence from unknown higher order constants.  Thus the initial scales need to be chosen carefully based on physical arguments.

For the purpose of logarithm resummation, the initial scales are usually chosen to minimize the logarithmic contributions at these scales, such that the fixed-order perturbation theory at the initial scales is already a good approximation. Since the logarithms are determined by the physical scale $Q$ as $\ln\frac{\mu^2}{Q^2}$,  an optimal choice is $\mu=Q$ so that the higher-order logarithms vanish. Based on this principle,
the hard scales in the Sudakov factors are determined by the external quark (antiquark) momentum $2|x\pm\xi|P_z$, corresponding to the logarithms $\ln\frac{\mu^2}{(2|x+\xi|P_z)^2}$ and $\ln\frac{\mu^2}{(2|x-\xi|P_z)^2}$, thus we can choose
\begin{align}
    \mu_{h_1}=2|x+\xi|P_z,\qquad& \qquad \mu_{h_2}=2|x-\xi|P_z.
\end{align}

On the other hand, the semi-hard scale in the jet function is more complicated. Naively, the explicit scale in its logarithm $\ln\frac{\mu^2}{(2|x-y|P_z)^2}$ suggests $\mu_i=2|x-y|P_z$,  depending on the momentum transfer carried by the emitted gluon. However, this scale depends on the variable $y$, which is a variable being integrated over in the matching. Thus for any $x$, there always exists a region $2|x-y|p\ll\Lambda_\text{QCD}$ during the integration, where the scale becomes non-perturbative. Then the method is numerically not implementable because it hits the Landau pole. There have been proposals to avoid the Landau pole with specific prescriptions~\cite{Catani:1996yz,Vogt:1999xa,Schoenleber:2022myb}, but these approaches introduce unphysical power corrections to the result~\cite{Beneke:1995pq}.
Here we adopt another approach to avoid the issue, which was initially proposed in the threshold resummation of DIS and Drell-Yan~\cite{Becher:2006mr,Becher:2007ty}, arguing that the actual semi-hard scale in the jet function should be determined after convoluting the jet function with a parton distribution functional form and integrating out the variable $y$, such that the logarithms in the results eventually only depend on the external physical scale $Q$ and the kinematic variable $x$. In our case, the scale will eventually depend on a combination of external momenta $xP_z$ and $\xi P_z$. In the threshold factorization of DIS, the resummed jet function is convoluted with the PDF $\phi(y)$ in a range $\xi\in[x,1]$~\cite{Becher:2006mr},
\begin{align}\label{eq:dis_log}
    F_{\text{DIS}}(x,Q^2)\propto \tilde{J}_z\left(\ln\frac{Q^2}{\mu_i^2}+\partial_\eta,\mu_i\right)\frac{e^{-\gamma_E\eta}}{\Gamma(\eta)}\int_x^1dy\frac{\phi_q(y,\mu_F)}{[(y/x-1)^{1-\eta}]_*},
\end{align}
and in the threshold region $\xi\to1$, $\phi(y)$ is approximated by a power-law function,
\begin{align}
    \phi_q(y)\sim(1-y)^b,
\end{align}
which simplifies the above integral to
\begin{align}
    F_{\text{DIS}}(x,Q^2)\propto \tilde{J}_z\left(\ln\frac{Q^2}{\mu_i^2}+\partial_\eta,\mu_i\right)\frac{e^{-\gamma_E\eta}\Gamma(b+1)}{\Gamma(\eta+b+1)}\left(\frac{1-x}{x}\right)^{\eta}.
\end{align}
When factoring out the $x$-dependence, the commutator $[\partial_\eta,\left(\frac{1-x}{x}\right)^{\eta}]=\left(\frac{1-x}{x}\right)\ln\left(\frac{1-x}{x}\right)$ introduces extra factors to the logarithms in the jet function,
\begin{align}
    F_{\text{DIS}}(x,Q^2)\propto \left(\frac{1-x}{x}\right)^{\eta}\tilde{J}_z\left(\ln\frac{(1-x)Q^2}{x\mu_i^2}+\partial_\eta,\mu_i\right)\frac{e^{-\gamma_E\eta}\Gamma(b+1)}{\Gamma(\eta+b+1)},
\end{align}
indicating that the physical scale is $\sqrt{\frac{1-x}{x}}Q$, corresponding to the invariant mass of the final state.

The argument is not directly applicable to the case of off-forward distributions, such as the GPD or DA, because the integral in Eq.~\eqref{eq:dis_log} is no longer limited to $[x,1]$, and a simple power-law cannot describe the lightcone distribution in all regions. Thus trying to obtain an analytical result of the integration to study the form of logarithm becomes impossible. Based on the same idea, we propose a different approach to analyze the form of logarithm after integration for more general cases.

Taking the factorization of quasi-GPD for an example, the convolution of jet function and the lightcone GPD is (as a general discussion, we ignore $\xi$- or $t$-dependence and only keep the $x$-dependence in the distributions $F$ and $\tilde{F}$)
\begin{align}
    \tilde{F}(x)\xrightarrow{}&\exp{\left[-2S_\Gamma(\mu_i,\mu)+a_J(\mu_i,\mu)\right]}\\
    &\times\tilde{J}_z\left(\ln{\frac{\mu^2}{4P_z^2}}-2\partial_\eta,\alpha(\mu_i)\right)\frac{\Gamma(1-\eta)e^{-\eta\gamma_E}}{\pi}\int_{-1}^{1} dy F(y)\left[\frac{\sin(\eta\pi/2)}{|y-x|^{1-\eta}}\right]_*. \nonumber
\end{align}
Note that the star function requires us to subtract $F(x)$ and its derivative up to order $\lfloor-\eta\rfloor$ at the singular point $y=x$, thus the integral should look like
\begin{align}
    \tilde{F}(x)\propto &\exp{\left[-2S_\Gamma(\mu_i,\mu)+a_J(\mu_i,\mu)\right]}\tilde{J}_z\left(\ln{\frac{\mu^2}{4P_z^2}}-2\partial_\eta,\alpha(\mu_i)\right)\frac{\Gamma(1-\eta)e^{-\eta\gamma_E}}{\pi}\sin(\eta\pi/2)\nonumber\\
    &\times\left[\int_{-1}^{1} dy \frac{F(y)-\sum^{\lfloor-\eta\rfloor}_{i=0} \frac{1}{i!}(y-x)^iF^{(i)}(x)}{|y-x|^{1-\eta}}\right.\nonumber\\
    &\quad\left.-\int_{1}^{\infty} dy \frac{\sum^{\lfloor-\eta\rfloor}_{i=0} \frac{1}{i!}(y-x)^iF^{(i)}(x)}{|y-x|^{1-\eta}}-\int_{\infty}^{-1} dy \frac{\sum^{\lfloor-\eta\rfloor}_{i=0} \frac{1}{i!}(y-x)^iF^{(i)}(x)}{|y-x|^{1-\eta}}\right],
\end{align}
where the two terms in the last line comes from the long-tail virtual contribution of the star function, labeled as $\tilde{F}_2(x)$. Physically, they correspond to residuals of the soft singularity cancellation between real and virtual contributions, which is the actual sources of the threshold logarithms in $\tilde{F}(x)$. 

First, we show that the first term in the square bracket, labeled as $\tilde{F}_1(x)$, does not contribute to threshold logarithm due to the cancellation between real and virtual contributions.
If $F(x)$ is a $C^{\infty}$ function in both $(a,x)$ and $(x,b)$, we can divide the integral into two parts, and Taylor expand $F(y)$ to get, 
\begin{align}\label{eq:scale_smooth1}
    \tilde{F}_1(x)&\propto\tilde{J}_z\left(\ln{\frac{\mu^2}{4P_z^2}}-2\partial_\eta,\alpha(\mu_i)\right)\int_{a}^{b} dy \frac{\sum^{\infty}_{n=1+\lfloor-\eta\rfloor} (y-x)^nF^{(n)}(x)}{n!|y-x|^{1-\eta}}
    \nonumber\\
    &\propto\tilde{J}_z\left(\ln{\frac{\mu^2}{4P_z^2}}-2\partial_\eta,\alpha(\mu_i)\right) \left[\int_x^{b} dy \sum^{\infty}_{n=1+\lfloor-\eta\rfloor} \frac{(y-x)^{n+\eta-1}}{n!}F^{(n)}(x)\right.\nonumber\\
    &\qquad+\left.\int_{a}^{x} dy \sum^{\infty}_{n=1+\lfloor-\eta\rfloor} (-1)^n\frac{(x-y)^{n+\eta-1}}{n!}F^{(n)}(x)\right] \nonumber\\
    &\propto \sum^{\infty}_{n=1+\lfloor-\eta\rfloor}\tilde{J}_z\left(\ln{\frac{\mu^2} {4P_z^2}}-2\partial_\eta,\alpha(\mu_i)\right)F^{(n)}(x)\left[\frac{(b-x)^{n+\eta}}{n!(n+\eta)} +(-1)^n \frac{{(x-a)}^{n+\eta}}{n!(n+\eta)} \right]\nonumber\\
    &\propto \sum^{\infty}_{n=1+\lfloor-\eta\rfloor} F^{(n)}(x) \frac{(b-x)^{n+\eta}}{n!(n+\eta)}\tilde{J}_z\left(\ln{\frac{\mu^2} {4(b-x)^2P_z^2}},\alpha(\mu_i)\right) \nonumber\\&\qquad + \sum^{\infty}_{n=1+\lfloor-\eta\rfloor}F^{(n)}(x)(-1)^n\frac{(x-a)^{n+\eta}}{n!(n+\eta)}\tilde{J}_z\left(\ln{\frac{\mu^2} {4(x-a)^2P_z^2}},\alpha(\mu_i)\right),
\end{align}
where the logarithm in the jet function eventually depends on the distance to the boundary, $|x-a|$ and $|x-b|$ , when convoluted with a $C^{\infty}$ distribution. However, note that $n+\eta\geq1+\eta+\lfloor-\eta\rfloor>0$, $(x-a)^{n+\eta}\ln(x-a)^m$ for any $m$ and $n$ in the series will be finite, so they are not logarithmically divergent. As a result, $\tilde{F}_1(x)$ is negligible when the threshold logarithm becomes important. On the other hand, the long-tail virtual contributions from the star function are,
\begin{align}\label{eq:scale_smooth2}
    \tilde{F}_2(x)&\propto\tilde{J}_z\left(\ln{\frac{\mu^2}{4P_z^2}}-2\partial_\eta,\alpha(\mu_i)\right)\left[\int_{b}^{\infty} dy \frac{\sum_{n=0}^{\lfloor-\eta\rfloor} (y-x)^nF^{(n)}(x)}{n!|y-x|^{1-\eta}}\right.
    \nonumber\\
    &\qquad+(-1)^n \left. \int_{-\infty}^{a} dy \frac{\sum_{n=0}^{\lfloor-\eta\rfloor} (y-x)^nF^{(n)}(x)}{n!|y-x|^{1-\eta}}\right]
    \nonumber\\
    &\propto \sum_{n=0}^{\lfloor-\eta\rfloor}\tilde{J}_z\left(\ln{\frac{\mu^2} {4P_z^2}}-2\partial_\eta,\alpha(\mu_i)\right)F^{(n)}(x)\left[\frac{(b-x)^{n+\eta}}{n!(n+\eta)} +(-1)^n \frac{{(x-a)}^{n+\eta}}{n!(n+\eta)} \right]\nonumber\\
    &\propto \sum_{n=0}^{\lfloor-\eta\rfloor} F^{(n)}(x) \frac{(b-x)^{n+\eta}}{n!(n+\eta)}\tilde{J}_z\left(\ln{\frac{\mu^2} {4(b-x)^2P_z^2}},\alpha(\mu_i)\right) \nonumber\\&
    \qquad + \sum_{n=0}^{\lfloor-\eta\rfloor}F^{(n)}(x)(-1)^n\frac{(x-a)^{n+\eta}}{n!(n+\eta)}\tilde{J}_z\left(\ln{\frac{\mu^2} {4(x-a)^2P_z^2}},\alpha(\mu_i)\right),
\end{align}
with $n+\eta\leq0$. Thus the logarithms from $\tilde{F}_2(x)$ are important and indeed the source of threshold logarithm that needs to be resummed. For a $C^{\infty}$ function $F(y)$ defined on $(a,b)$, the threshold logarithms generated from the convolution of the kernel $\mathcal{C}(x,y)$ on $F(y)$ are $\ln{\frac{\mu^2} {4(x-a)^2P_z^2}}$ and $\ln{\frac{\mu^2} {4(x-b)^2P_z^2}}$.

More generally, when the function is defined on $y\in(c,d)$, and is not $C^{\infty}$ at $y=a$ and $y=b$  with $d>b>a>c$, we can redefine a piece-wise function as a combination of $C^{\infty}$ functions:
\begin{align}
    F(y)=&\left\{\begin{matrix}
        G_1(y),& c<y<a \\
        G_1(y)-(G_1(y)-F(y)),& a<y<x\\
        G_2(y)-(G_2(y)-F(y)),& x<y<b\\
        G_2(y),& b<y<d
    \end{matrix}\right.,
    \\
    =&G_1(y)\Theta(c<y<x)+G_2(y)\Theta(x<y<d) \nonumber\\
    &+(F(y)-G_1(y))\Theta(a<y<x) +(F(y)-G_2(y))\Theta(x<y<b) ,
\end{align}
where $G_1(x)$ and $G_2(x)$ are smooth continuations of $F(x)$ from $(c,a)$ and $(b,d)$ to $(c,x)$ and $(x,d)$. Then the integral can be decomposed into four parts with a smooth integrand in each region: the convolution with $G_1$ on $(c,x)$;  the convolution with $(G_1-F)$ on $(a,x)$; the convolution with $(G_2-F)$ on $(x,b)$;  and the convolution with $G_2$ on $(x,d)$. Following the same derivation as Eq.~\eqref{eq:scale_smooth2} and Taylor expand the functions at $x$, we can get four different logarithms, $\ln{\frac{\mu^2} {4(x-c)^2P_z^2}}$, $\ln{\frac{\mu^2} {4(x-a)^2P_z^2}}$, $\ln{\frac{\mu^2} {4(b-x)^2P_z^2}}$, and $\ln{\frac{\mu^2} {4(d-x)^2P_z^2}}$. It can be further generalized to functions with any number of non-smooth points, where each non-smooth point $x_i$ in the integral induces a logarithm of $\ln{\frac{\mu^2} {4(x-x_i)^2P_z^2}}$, corresponding to a physical scale $2|x-x_i|P_z$.
However, not all of them are important in the perturbation series, because the integrand is always enhanced by the gluon momentum with a factor of $\frac{1}{|x-y|^{-n+1-\eta}}$, resulting in an enhancement in the final result $\frac{1}{|x-x_i|^{-n-\eta}}$ with  $n+\eta \leq0$, which is more divergent than  the logarithms, thus will filter out the nearest $|x-x_i|$ contribution. Eventually, only the nearest non-smooth point $y=x_0$ to $x$ induces the most important logarithmic contribution.

Thus we can define the semi-hard scale as a profile function that depends on the distance to the nearest non-smooth point. For the case of GPD, the non-smooth points are $\pm1$ and $\pm\xi$, corresponding to a profile function:
\begin{align}
    \mu_i=2\min[|1-x|,|1+x|,|\xi-x|,|\xi+x|]P_z.
\end{align}
In the ERBL region $|x|<\xi$, the two scales $2(\xi\pm x)P_z$ are the same as the two hard scales in the Sudakov factor, which is also the case for DA~\cite{Baker:2024zcd,Cloet:2024vbv}.
In the limit of $\xi\to0$, where the quasi-GPD factorization becomes the same as the quasi-PDF factorization, the scale $2xP_z$ is the hard scale corresponding to quark momentum, and the scale $2(1- |x|)P_z$ when $x\to\pm1$ are the threshold semi-hard scales~\cite{Ji:2023pba,Ji:2024hit}.

\subsection{Resummation of the full matching kernel}
The resummation of the full matching kernel includes two steps: evaluating the matching kernel at a specific physical scale where all the logarithms are as small as possible; then resumming the remaining logarithms in the threshold limit.

The first step requires us to find a proper scale $\mu_h$, where the matching kernel is obtained through a full evolution,
\begin{align}
    \mathcal{C}(x,y,\xi,P_z,\mu)=\mathcal{C}(x,y,\xi,P_z,\mu_h)\exp[\int_\mu^{\mu_h}\hat{\mathcal{V}}(\mu)d\ln\mu^2],
\end{align}
where $\hat{\mathcal{V}}$ is the GPD evolution kernel in Eq.~\eqref{eq:gpd_evo}, such that
\begin{align}
    \tilde{F}(x,\xi,P_z,t)=\int dy\mathcal{C}(x,y,\xi,P_z,\mu)F(y,\xi,\mu,t)=\int dy\mathcal{C}(x,y,\xi,P_z,\mu_h)F(y,\xi,\mu_h,t).
\end{align}
In the ERBL region $|x|<\xi$, the hadron is emitting a quark-antiquark pair with momentum fraction $\xi \pm x$. Thus it looks like emitting a meson of total momentum $2\xi P_z$, and it is natural to choose $\mu_h=2\xi P_z$ in the first step. In the DGLAP region $|x|>\xi$, we propose to evaluate the matching kernel at $\mu_h=2|x|P_z$ in order to enable the resummation in the threshold limit for any $\xi$ value, as shown in Sec.~\ref{Sec:Methodology}.

The second step includes multiplicatively subtracting the fixed-order threshold terms in the full matching kernel, then adding back the resummed version of the threshold terms~\cite{Baker:2024zcd,Cloet:2024vbv},
\begin{align}\label{eq:resum_full}
    \mathcal{C}_\text{TR}(\mu_h)=JH_\text{TR}(\mu_h,\mu_i,\mu_{h_1},\mu_{h_2})\otimes JH^{-1}_\text{NLO}(\mu_h)\otimes\mathcal{C}_\text{NLO}(\mu_h),
\end{align}
where $JH_\text{TR}(\mu_h,\mu_i,\mu_{h_1},\mu_{h_2})$ is the resummed kernel at scale $\mu_h$ in the threshold limit,
with semi-hard scale $\mu_i$ in the jet function, and hard scale $\mu_{h_{1,2}}$ in the Sudakov factors,
\begin{align}\label{eq:jh}
    JH_\text{TR}(\mu_h,\mu_i,\mu_{h_1},\mu_{h_2})=J(\mu_h,\mu_i)\otimes H(\mu_h,\mu_{h_1},\mu_{h_2}),
\end{align}
and $JH^{-1}_\text{NLO}(\mu_h)=JH_\text{TR}(\mu_h,\mu_h,\mu_h,\mu_h)$ is the fixed-order kernel in the threshold limit. Since the factorization is multiplicative in coordinate space, it is equivalent to write the momentum space version of $JH_\text{TR}$ in a different order,
\begin{align}\label{eq:hj}
    JH_\text{TR}(\mu_h,\mu_i,\mu_{h_1},\mu_{h_2})=H(\mu_h,\mu_{h_1},\mu_{h_2})\otimes J(\mu_h,\mu_i),
\end{align}
which differs from Eq.~\eqref{eq:jh} by higher order corrections of $\mathcal{O}(|x-y|^0)$ in the threshold expansion. Such a difference could be included as a systematic error of the calculation.

Combining the above two steps, the full matching kernel is resummed as
\begin{align}
    \mathcal{C}_\text{TR}(\mu)=JH_\text{TR}(\mu_h,\mu_i,\mu_{h_1},\mu_{h_2})\otimes JH^{-1}_\text{NLO}(\mu_h)\otimes\mathcal{C}_\text{NLO}(\mu_h)\otimes\exp[\int_\mu^{\mu_h}\hat{\mathcal{V}}(\mu')d\ln\mu'^2],
\end{align}
with the scales chosen as
\begin{align}\label{eq:physical_scales}
    \mu_h=2\max[|x|,\xi]P_z,\qquad \mu_{h_1}=2|x+\xi|P_z,\qquad
    \nonumber\\
    \mu_{h_2}=2|x-\xi|P_z,\qquad \mu_i=2\min[|x\pm\xi|,|1\pm x|]P_z.
\end{align}


\subsection{Resumming logarithm with leading power accuracy}
Ideally, the factorization formula of quasi-GPD in Eq.~\eqref{eq:factorization} has power corrections starting at quadratic order in $\Lambda_\text{QCD}/p_z$ with the physical scale $p_z=|x\pm\xi|P_z$ or $|1\pm x|P_z$. However, the quasi-GPD $\tilde{F}(x,\xi,P_z)$ renormalized in the hybrid scheme contains a linear renormalon from the spatial Wilson line~\cite{Beneke:1992ea,Beneke:1992ch,Beneke:1998ui,Bauer:2011ws,Bali:2013pla,Bali:2014fea,Bali:2014sja,Ayala:2019uaw,Ayala:2020pxq}. The same linear renormalon exists in the matching kernel $\mathcal{C}(x,y,\xi,P_z/\mu)$, corresponding to the factorially growing coefficients in the perturbative expansion~\cite{Braun:2018brg}. A naive inverse convolution of them will in principle introduce a linear power correction $\mathcal{O}\left(\frac{\Lambda_\text{QCD}}{p_z}\right)$ to the lightcone GPD $F(x,\xi,\mu)$. In the threshold factorization, this renormalon goes with the spatial Wilson line into the jet function. Meanwhile, another linear renormalon is introduced to the phase of the Sudakov factors due to the threshold factorization~\cite{Liu:2023onm}. Although the $JH$ and $JH^{-1}$ in Eq.~\eqref{eq:resum_full} contain opposite renormalons that cancel order by order, the renormalon in the fixed-order inverse kernel $JH^{-1}_\text{NLO}$ does not automatically cancel the one in the resummed kernel $JH_\text{TR}$ because of different higher-order logarithm terms. Thus, without renormalon regularization, the resummation of logarithms in the matching kernel will still introduce linear power corrections to the lightcone GPD.

To remove power corrections introduced by the renormalon series, there are two types of approaches in general. One is to define the sum of the factorially divergent series with known asymptotic behavior in certain regularization scheme, e.g., integrating along a fixed path on the Borel plane, which is regularization-scheme dependent~\cite{Pineda:2001zq,Bali:2014sja,Ayala:2019uaw}. Another type is to subtract the divergent part from the series and work with the remaining finite renormalon-free part, such as the ``R-evolution'' method~\cite{Hoang:2008yj,Hoang:2009yr}, which is subtraction-scheme dependent. In either case, a non-perturbative quantity is defined to absorb the scheme dependence and the renormalon ambiguity. A systematic method to achieve the leading power accuracy following the first strategy has been proposed recently for the LaMET calculations~\cite{Zhang:2023bxs}, and has been applied to the calculation of GPD without resumming the logarithms~\cite{Holligan:2023jqh}. The idea is to regularize all renormalons in the same scheme, such that the ambiguities corresponding to the same renormalon cancel in the convolution. The linear renormalon in the matching kernel or the Wilson coefficients is regularized by the LRR, which defines the scheme $\tau$ as the PV prescription when resumming the asymptotic series in the Borel plane. This approach ensures that all perturbation series containing the same renormalon is regularized in the same $\tau$-scheme. On the other hand, the renormalon in the renormalized lattice quasi-GPD is originally obtained in an different scheme $\tau'$, depending on how the mass counterterm $\delta m(a,\tau')$ in Eq.~\eqref{eq:hybrid_scheme} is extracted.
It is then converted to the same $\tau$ scheme by introducing a scheme-dependent non-perturbative parameter $m_0(\tau,\tau')$ to the  renormalization constant, $e^{\delta m(a,\tau')|z|} \to e^{(\delta m(a,\tau')+m_0(\tau,\tau'))|z|}$, which is extracted by matching the renormalized lattice data at $P_z=0$ in scheme $\tau'$ with LRR-improved perturbation theory at short distance in scheme $\tau$~\cite{Zhang:2023bxs}. 

The same method could be applied to our resummation formalism. Once we regularize all the linear renormalons in the same scheme, the ambiguities can  cancel between $H^{-1}_\text{NLO}$ and $H_\text{TR}$, as well as between $J^{-1}_\text{NLO}$ and $J_\text{TR}$. 
The coordinate space correction to the jet function is the same as that in the Wilson coefficient $C_0$~\cite{Zhang:2023bxs},
\begin{align}
    \tilde{J}_z^\text{LRR}(z,\mu)=\tilde{J}_z+|z|\mu\left(R(\alpha_s)_\text{PV}-\sum_{i=0}^n r_i\alpha_s^{i+1}\right),
\end{align}
where $r_i$ is the coefficient of the asymptotic series,
\begin{align}
    r_i = N_m \left(\frac{\beta_0}{2\pi}\right)^i
    \frac{\Gamma(n+1+b)}{\Gamma(1+b)}
    \left[1+\frac{c_1b}{b+i} +...\right],
\end{align} 
with $b=\beta_1/2\beta_0^2$, $c_1=(\beta_1^2-\beta_0\beta_2)/(4b\beta^4_0)$, and $N_m(n_f=3) =0.575$ in the $\overline{\text{MS}}$ scheme.  $R_\text{PV}$ is the resummation of the series $r_i\alpha_s^{i+1}$ with the PV prescription,
\begin{align}
\label{eq:borel_integral}
    &R(\alpha_s)_\text{PV}= N_m\frac{4\pi}{\beta_0}   \int_{0, \text{PV}}^{\infty}du   e^{-\frac{4\pi u}{\alpha_s\beta_0}}  \frac{1}{(1-2u)^{1+b}}\bigg(1+c_1(1-2u)+...\bigg).
\end{align}
Since the threshold resummation of $J_\text{TR}$ is defined in coordinate space first and then Fourier transformed to the momentum space, we perform the same operation on the correction term,
\begin{align}
    J^\text{LRR}_\text{TR}(\mu)=&J_\text{TR}+\int \frac{dzP_z}{2\pi}e^{i(x-y)zP_z}|z|^{1-2a_\Gamma(\mu_i,\mu)}e^{-2S_\Gamma(\mu_i,\mu)+a_J(\mu_i,\mu)}\mu_i\left(\frac{\mu_ie^{\gamma_E}}{2}\right)^{-2a_\Gamma(\mu_i,\mu)}\nonumber\\
    &\times \left(R(\alpha_s(\mu_i)_\text{PV}-\sum_{i=0}^n r_i\alpha_s^{i+1}(\mu_i)\right),
\end{align}
where the Fourier transformation of $|z|^{1-2a_\Gamma(\mu_i,\mu)}$ is obtained using the same $\epsilon_m$ regularization as in Ref.~\cite{Zhang:2023bxs},
\begin{align}
    \int \frac{dz}{2\pi}e^{i(x-y)zP_z}|z|^{1-2a_\Gamma}e^{-|z|\epsilon_m}=\frac{\Gamma(2-a_\Gamma)}{\pi}\Re[(\epsilon_m-i|x-y|P_z)^{-2-a_\Gamma}].
\end{align}

Note that although the resummed renormalon series $|z|\mu R(\alpha_s)_\text{PV}$ explicitly depends on the scale $\mu$, the renormalon ambiguity~\cite{Beneke:1998ui}, 
\begin{align}
    \Res[|z|\mu R(\alpha_s)]\propto |z|N_m\Lambda_\text{QCD},
\end{align}
is $\mu$-independent. This allows the cancellation between the renormalon ambiguities in $J_\text{TR}(\mu)$ and $J^{-1}_\text{NLO}(\mu)$ with renormalon regularized in the same scheme  but at different scales. Same cancellation works for the renormalons in the Sudakov factors below.

The LRR of the Sudakov factor contains two parts: the LRR correction to the hybrid scheme correction $\Delta|H|(p_z,\mu)$ in Eq.~\eqref{eq:hybrid_corr},
\begin{align}\label{eq:hybrid_corr_lrr}
    \Delta|H|^\text{LRR}(p_z,\mu_h)=\Delta|H|(p_z,\mu_h)-\frac{z_s\mu_h}{2}\left(R(\alpha_s(\mu_h)_\text{PV}-\sum_{i=0}^n r_i\alpha_s^{i+1}(\mu_h)\right),
\end{align}
and the LRR of the phase factor $A(p_z,\mu)$~\cite{Ji:2024hit} (note that the phase factor in GPD is half of the phase factor in PDF because the Sudakov factor is factorized into two parts),
\begin{align}
    A^\text{LRR}(p_z,\mu_h)=A(p_z,\mu_h)+\frac{\mu_h}{|p_z|}\left(R(\alpha_s(\mu_h)_\text{PV}-\sum_{i=0}^n r_i\alpha_s^{i+1}(\mu_h)\right).
\end{align}
The LRR in the perturbative matching defines the quasi-GPD in a specific scheme of renormalon regularization. Thus the renormalized lattice quasi-GPD also depends on the renormalon regularization, and this scheme dependence is only canceled when matched to lightcone GPD with the matching kernel regularized in the same scheme.
Once all the renormalons are regularized in the same schemed defined by the PV prescription in LRR, the resummation of logarithms will not introduce extra linear power correction, and the power accuracy of the factorization formalism Eq.~\eqref{eq:factorization} is achieved. 

\section{Numerical Tests} 
\label{Sec:Numerics}

With the formula derived above, we now test the method on a GPD model.
Here, we choose the double-distribution GPD model used in Refs.~\cite{Radyushkin:1997ki,Belitsky:2005qn},
\begin{align}
F(x,\xi,t) &= \theta(x+\xi)\frac{2+\lambda}{4\xi^3} \left(\frac{x+\xi}{1+\xi}\right)^\lambda \left[\xi^2-x+\lambda\xi(1-x)\right]
    \\
    &{}-\theta(x-\xi)\frac{2+\lambda}{4\xi^3} \left(\frac{x-\xi}{1-\xi}\right)^\lambda \left[\xi^2-x-\lambda\xi(1-x)\right]
\end{align}
with $\xi=0.5$ and $\lambda=1.5$, such that both the ERBL and DGLAP regions are large enough for us to examine the resummation effect.
Meanwhile, the lightcone distribution is in general factorization-scale dependent.
Thus, we need to specify the value of $\mu$ when using this model, and the distribution at different $\mu$ values can be connected through the evolution of GPD given in Eq.~\eqref{eq:gpd_evo}.
As the physical scales in Eq.~\eqref{eq:physical_scales} could become non-perturbative when $x$ is close to the boundaries $\pm\xi$ and $\pm 1$, the range we could calculate in the LaMET framework is roughly bounded to be
\begin{equation}
    x\in \mathcal{X}\equiv[-1+x_0,-\xi-x_0]\cup[-\xi+x_0,\xi-x_0]\cup[\xi+x_0,1-x_0],
\end{equation}
with the cutoff $x_0$ given by $\sim \Lambda_\text{QCD}/P_z$, as demonstrated in the calculations of PDF~\cite{Su:2022fiu,Ji:2024hit} and DA~\cite{Baker:2024zcd,Cloet:2024vbv}.
In practice, lattice calculations involve hadron momenta $P_z\approx 2$~GeV, then the truncation will be around $x_0\approx0.2$, as we will show below. It will not be an issue if we only need the matching to the quasi-GPD, or only the inverse matching from known quasi-GPD to lightcone GPD. In this work, however, we only test on a lightcone GPD model, and want to obtain the quasi-GPD followed by an inverse matching on the interpolated function to reproduce the original GPD model. $x_0\approx 0.2$  creates a too large gap that makes the interpolation meaningless. So here we choose $\mu=P_z=4$~GeV to further suppress $x_0$, such that an interpolation of the quasi-GPD results is possible.
Then, we can apply an inverse matching to the interpolated quasi-GPD and compare with the original GPD, which is a good test of the self-consistency of our resummation method.

\begin{figure}
    \centering
    \includegraphics[width=0.5\linewidth]{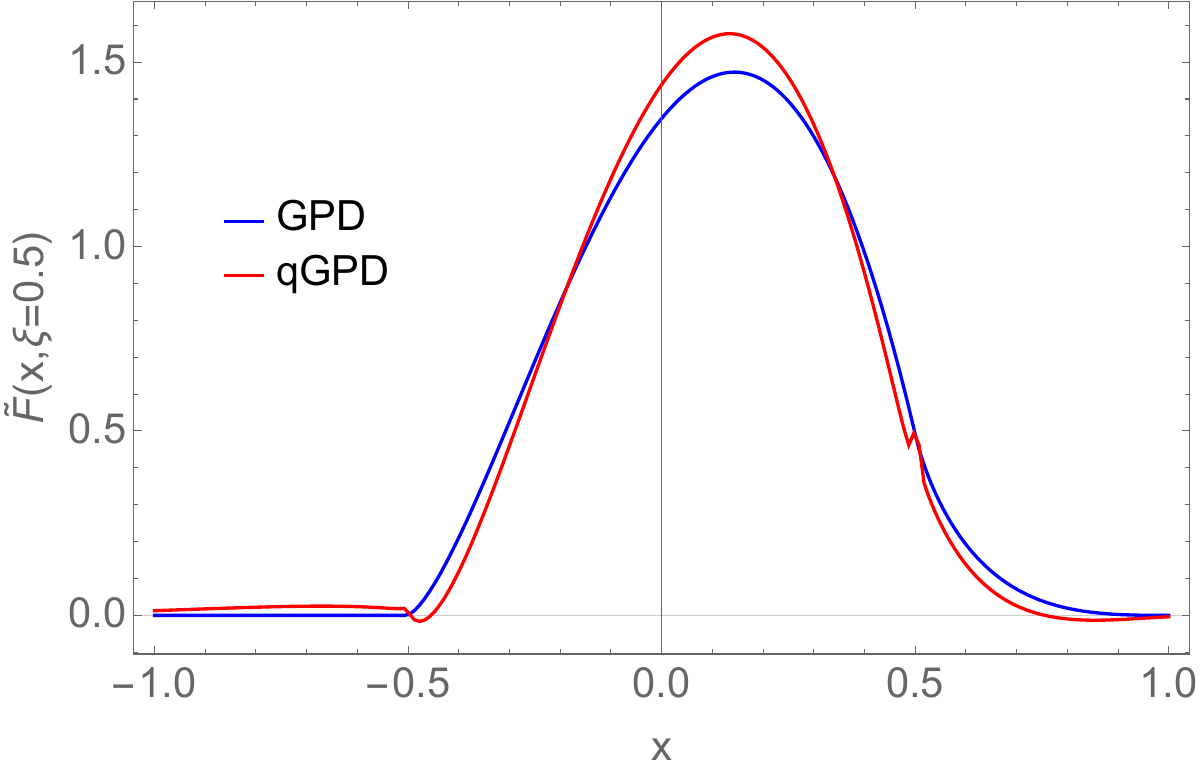}
    \caption{The GPD model at $\mu=4$~GeV and the corresponding quasi-GPD at $P_z=4$~GeV with NLO matching. }
    \label{fig:fo_matching}
\end{figure}

With this setup, a direct fixed-order matching $\mathcal{C}_\text{NLO}(x,y,\xi,P_z,\mu)$ could be convoluted with the GPD model $F(y,\xi,\mu,t)$ to obtain the NLO quasi-GPD $\tilde{F}(x,\xi,P_z,t)$ in the hybrid scheme with $z_s=0.2$~fm (a typical scale in lattice calculations that satisfies $a\ll z_s\ll \Lambda^{-1}_{\rm QCD}$), as shown in Fig.~\ref{fig:fo_matching}.
Since we implement the matching as matrix product, the process is reversible, and an inverse matching matrix $\mathcal{C}^{-1}_\text{NLO}(x,y,\xi,P_z,\mu)$ applied to the quasi-GPD $\tilde{F}(x,\xi,P_z,t)$ exactly reproduces the original GPD model.

\begin{figure}
    \centering
    \includegraphics[width=0.45\linewidth]{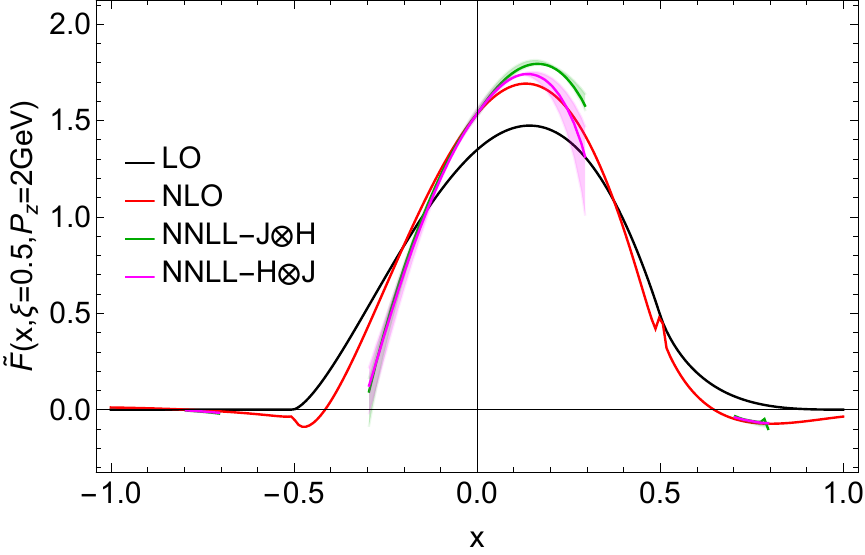}
    \includegraphics[width=0.45\linewidth]{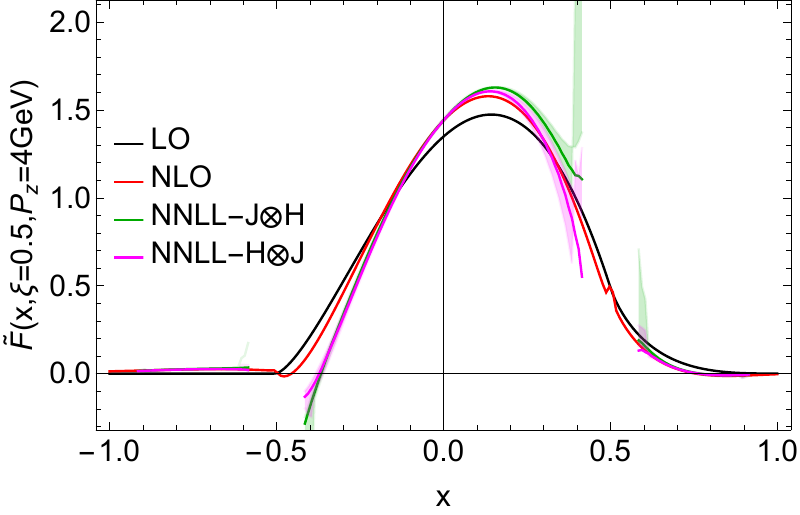}
    \caption{The perturbatively matched quasi-GPD with NLO and resummed NNLL matching kernels at $P_z=2$~GeV (left) and $P_z=4$~GeV (right) for $\xi=0.5$. The bands represents scale variation in the initial scale choice of the resummation. Qualitatively, the matching effects with different physical scales are similar. At larger momentum, we are able to calculate a larger range of $x$, and the difference between the two methods are smaller because the higher-order threshold terms are also more suppressed.}
    \label{fig:nnll_matching}
\end{figure}

\begin{figure}
    \centering
    \includegraphics[width=0.49\linewidth]{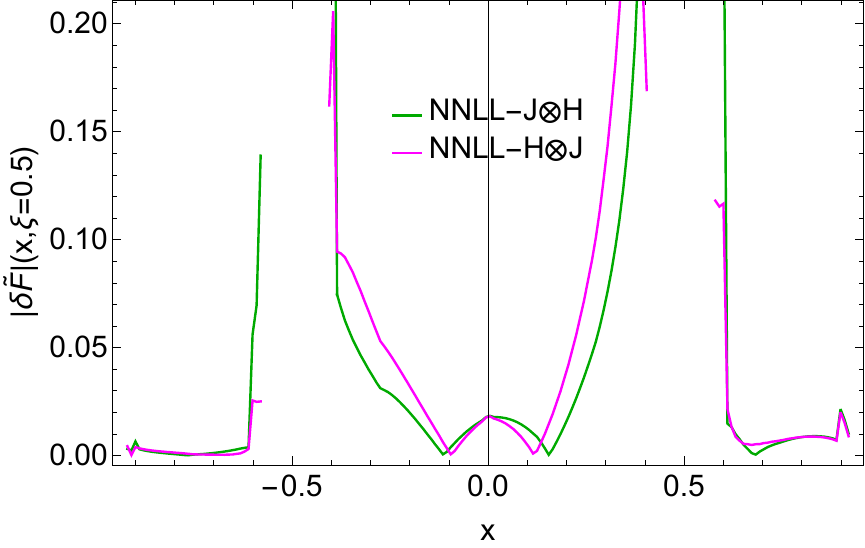}
    \includegraphics[width=0.49\linewidth]{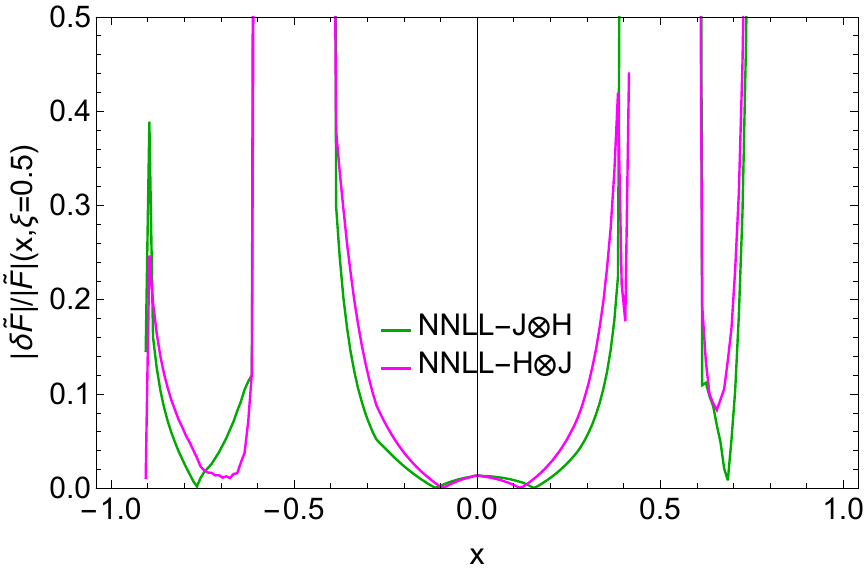}
    \caption{The absolute (left) and relative (right) scale variation of quasi-GPD as a function of $x$. The divergence in the scale variation (except for the relative uncertainty near $x=0.8$ because $\tilde{F}(0.8)\approx 0$) is a clear sign of entering the non-perturbative region, where we cannot obtain reliable results from perturbative matching.}
    \label{fig:scale_var}
\end{figure}

\begin{figure}
    \centering
    \includegraphics[width=0.5\linewidth]{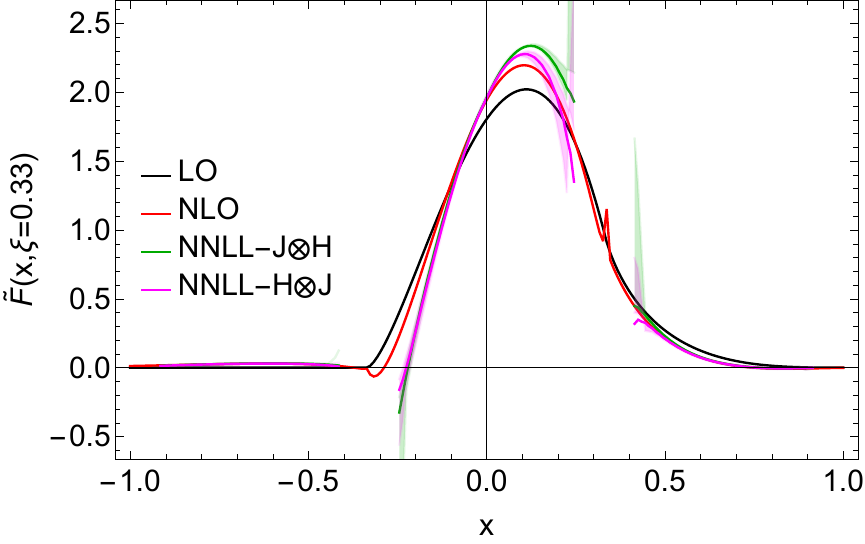}
    \caption{The perturbatively matched quasi-GPD with NLO and resummed NNLL matching kernels at $P_z=4$~GeV for $\xi=1/3$. The bands represents scale variation in the initial scale choice of the resummation. We observe a similar truncation $x_0\sim0.1$ where the physical scales start to become non-perturbative.}
    \label{fig:different_xi}
\end{figure}

Once we resum the matching kernel, the calculation explicitly diverges in the non-perturbative regions.
Then, we get three pieces of quasi-GPD from the calculation, as shown in Fig.~\ref{fig:nnll_matching}.
Here we show both $P_z=2$~GeV and $P_z=4$~GeV for comparison (note that the model is always defined at $\mu=P_z$ for simplicity). Qualitatively, the matching effects at different $P_z$ are similar, but they are smaller at larger $P_z$ due to smaller coupling $\alpha_s$ at larger physical scales. 
Note that there is a slight difference in the two different orders of the convolution $JH= J\otimes H$ or $H\otimes J$, which comes from the higher-order threshold terms in $|x-y|$ and should be treated as a systematic error. At larger momentum, we are able to calculate a larger range of $x$, and the difference between the two methods is smaller because the higher-order threshold terms are more suppressed. 
To examine higher-order effects and test the stability of the perturbative calculation, we vary the physical scales by a factor of $\sqrt{2}$, i.e., $\mu_{i/h/h_{1}/h_{2}}\to c\mu_{i/h/h_{1}/h_{2}}$ with $c=\{\sqrt{2},1,\sqrt{2}^{-1}\}$.
Then, the uncertainties from scale variation is an estimate of potential higher-order corrections.
Figure~\ref{fig:scale_var} shows the scale variation as a function of $x$.
At $x=0$, all the relevant physical scales in the system are identical, $\mu_i=\mu_h=\mu_{h_1}=\mu_{h_2}=cP_z$. In this case, the two implementations are exactly the same, and there is no threshold resummation effect. However, there is still a non-vanishing scale variation due to the overall evolution effect of GPD from scale $c\mu_h$ to $\mu$. We also observe a diverging trend when $|x\pm \xi|<0.1$ and $|x\pm 1|<0.1$. The divergence of scale variation is a sign of the breakdown of perturbation theory, thus setting a bound on $x$, only within which could we obtain reliable results from perturbative matching. 
{According to the physical implication of resummation, the truncation $x_0\approx 0.1$ is determined by the condition that one of the physical scales in the system drops to $2x_0P_z$ and becomes non-perturbative such that $\alpha_s(2x_0P_z)\sim1$,
which is not sensitive to $\xi$. A similar $x_0$ is observed in a different test with $\xi=1/3$, as shown in Fig.~\ref{fig:different_xi}.
This prevents us from extracting the $x$-dependence near $x\to\pm\xi$ and $x\to\{0,1\}$. Nevertheless, it could in principle be improved with larger $P_z$ in future lattice calculations, especially with the recent developments~\cite{Bali:2016lva,Zhang:2025hyo} that improve the precision of measuring boosted hadrons. Meanwhile, as the experiments provide useful constraints on the $x$-dependence near $x\to\pm\xi$, the LaMET calculations provide $x$-dependence in $x\in\mathcal{X}$, and lattice calculations of moments provide global information of GPDs, we can hopefully fully understand the GPDs in future by combining all the complementary information~\cite{Ji:2022ezo}.}

\begin{figure}
    \centering
    \includegraphics[width=0.5\linewidth]{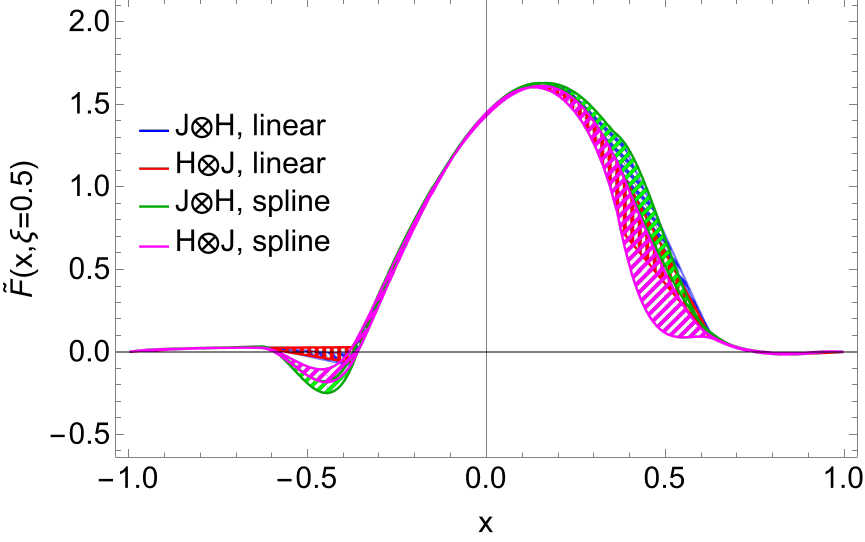}
    \caption{Different interpolations of quasi-GPD function with resummed matching kernels. The large uncertainty in the interpolated region can be regarded as an estimate of power corrections and non-perturbative effects there.}
    \label{fig:interpolation}
\end{figure}

Since we do not have a full quasi-GPD calculated with the resummed formalism, we are unable to implement an inverse matching directly. However, considering that the $x$-dependent quasi-GPD calculated on lattice is usually a continuously smooth function, this should not be an issue for the practical implementation of our formalism. Therefore, for the purpose of model demonstration, we can interpolate the current quasi-GPD calculated on $\mathcal{X}$ to extend its domain to the full $x\in[-1,1]$, including the regions where matching breaks down. In order to check how sensitive our final result is to the unknown interpolated region, we try different interpolation orders for comparison.  We choose a simple linear interpolation and a spline interpolation and show the comparison in Fig.~\ref{fig:interpolation}. The latter guarantees that the second order derivative is continuous, making the curve ``look smooth''. The uncertainty in the interpolated region $x\to\pm\xi$ is significantly larger than other regions, and could be a rough estimate of these unknown and uncontrollable effects. Note that for the twist-2 contribution we do not need to know the quasi-distribution in $|x|>1$ to calculate lightcone GPD, as long as we only invert the matching (sub-)matrix defined in $|x|,|y|<1$, so it is ignored in our interpolation. In practical lattice calculations, it is still recommended to convolute the inverse matching with a full range of quasi-distribution in $[-\infty,\infty]$.

\begin{figure}
    \centering
    \includegraphics[width=0.5\linewidth]{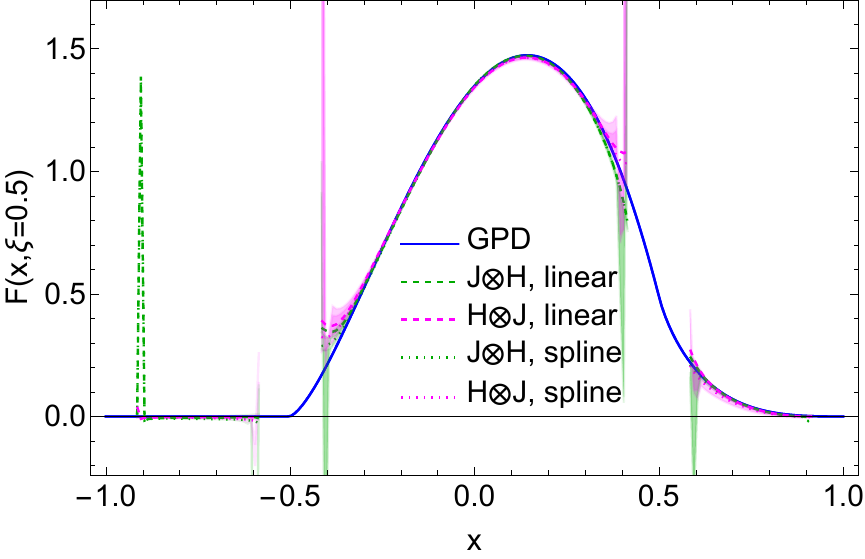}
    \caption{Lightcone GPD results obtained from applying the resummed inverse matching kernels to the interpolated quasi-GPD results calculated from resummed matching kernels in the hybrid scheme. Different approaches and interpolations all reproduce the original GPD model consistently within the perturbative region.}
    \label{fig:inverse_matching}
\end{figure}

Then we implement the resummed inverse matching kernel to the above interpolated quasi-GPD functions. If we ignore the interpolation step, the resulting lightcone GPD from our numerical approach is related to the original GPD through (the LRR correction is implied in the matching kernel and threshold terms):
\begin{align}
        F(x,\mu,&\xi,t)=\left(\hat{\mathcal{V}}(\mu_h,\mu)\otimes \mathcal{C}^{-1}_\text{NLO}(\mu_h)\otimes JH_\text{NLO}(\mu_h)\otimes JH^{-1}_\text{TR}(\mu_h,\mu_i,\mu_{h_1},\mu_{h_2})\right)_{x,y}\\
        &\cdot\left(  JH_\text{TR}(\mu'_h,\mu'_i,\mu'_{h_1},\mu'_{h_2})\otimes JH_\text{NLO}^{-1}(\mu'_h)\otimes\mathcal{C}_\text{NLO}(\mu'_h)\otimes \hat{\mathcal{V}}(\mu,\mu'_h)\right)_{y,z}F(z,\mu,\xi,t),\nonumber
\end{align}
where all scales without the prime symbol ``$'$'' are determined by the variable $x$, and all scales with the prime symbol ``$'$'' depend on variable $y$. Clearly, due to the different scale choices, the resummed inverse matching kernels in the first line is mathematically different from the inverse of the second line, and they in principle do not completely cancel each other. However, after implementing the resummed inverse matching, we do find the quasi-GPD is converted back to the original GPD curve in $x\in\mathcal{X}$ with high accuracy, as shown in Fig.~\ref{fig:inverse_matching}. Moreover, the two different implementations $JH= J\otimes H$ or $H\otimes J$ generate almost identical results before entering the non-perturbative region. It is a sign that our method is self-consistent, especially at large momentum.

More interestingly, regardless of how the incalculable non-perturbative region is interpolated, the inversely matched results are consistent.
Our results suggest that the unknown systematics in non-perturbative regions do not introduce a noticeable uncertainty in perturbative region after the inverse matching in LaMET. 

{
To demonstrate the renormalization-scheme independence of the method, we also test the same procedure in the $\overline{\rm MS}$ scheme. Then, none of Eq.~\eqref{eq:hybrid_matching}, Eq.~\eqref{eq:hybrid_corr}, or Eq.~\eqref{eq:hybrid_corr_lrr} is needed. In the $\overline{\rm MS}$ scheme, there is no current conservation in the perturbative matching step due to the existence of the extra term proportional to $\alpha_s\delta(x-y)$ in Eq.~\eqref{fig:fo_matching}. As a result, the quasi-GPD in the $\overline{\rm MS}$ scheme is noticeably higher than the hybrid scheme, but the enhancement will be compensated by the same suppression in the inverse matching, as shown in Fig.~\ref{fig:msbar}. The method is again self-consistent when implemented in the $\overline{\rm MS}$ scheme. The cut-off is also $x_0\approx 0.1$, because the $\Lambda_{\rm QCD}$ and $\alpha_s$ are defined in the same way in these two schemes.
\begin{figure}[t]
    \centering
    \includegraphics[width=0.49\linewidth]{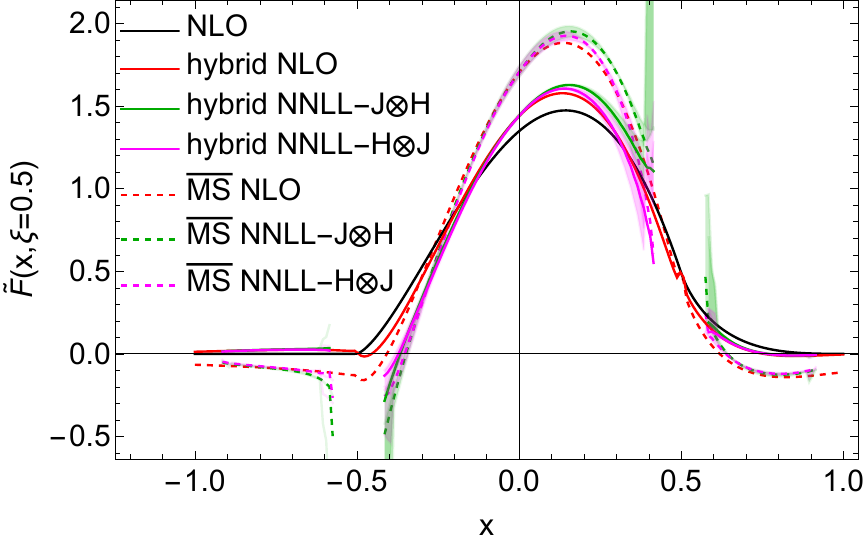}
    \includegraphics[width=0.49\linewidth]{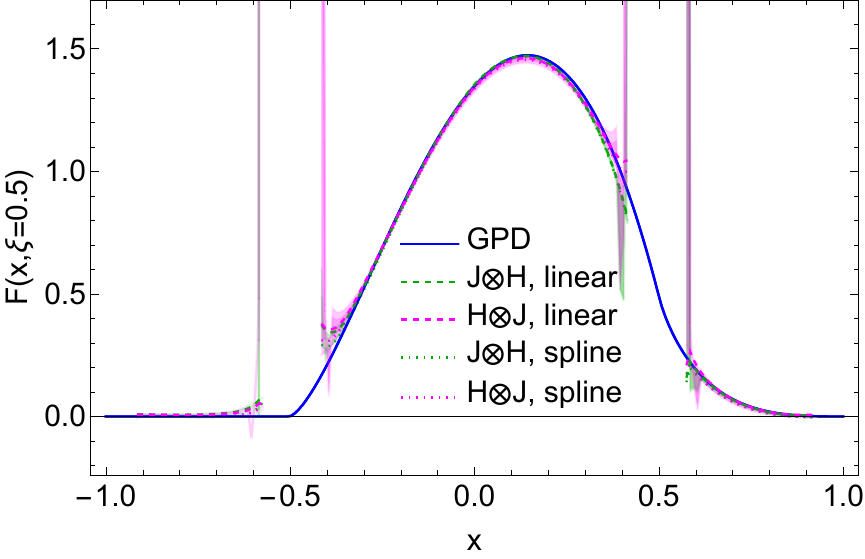}
    \caption{The resummed GPD matching comparison between the hybrid and the $\overline{\rm MS}$ schemes (left), and inverse matching after an interpolation in the $\overline{\rm MS}$ scheme (right) for $\xi=0.5$. Although the quasi-GPD in the $\overline{\rm MS}$ scheme is much higher, the inverse matching consistently reproduces the original GPD model.}
    \label{fig:msbar}
\end{figure}
}

\section{Conclusion}\label{Sec:Conclusion}
In this work, we use SCET to derive the threshold factorization for the quasi-GPD in the soft-gluon-emission limit, which can reduce to the quasi-PDF and quasi-DA cases with their corresponding external states. Based on the factorization formula, we propose a method to resum the large logarithms related to soft partonic momenta in the perturbative matching of quasi-GPD. There are three different scales in the matching kernel, $2|x\pm\xi|P_z$ from the quark (antiquark) momentum and $2|x-y|P_z$ from the gluon momentum, which makes it difficult to resum directly. We demonstrate that these logarithms are only important in the threshold limit in the ERBL region for any $\mu\sim P_z$, and in the DGLAP region when setting $\mu=2xP_z$. Based on this finding, we further factorize the matching kernel in the threshold limit, then resum the three different logarithms independently. We discuss the choice of initial scales in the solution of RG equations, and derive the correction to the full matching kernel. 
By applying the resummed matching kernel to a GPD model, we demonstrate that the reliable range of LaMET calculation is $\mathcal{X}=[-1+x_0,-\xi-x_0]\cup[-\xi+x_0,\xi-x_0]\cup[\xi+x_0,1-x_0]$ in the quasi-GPD, where the cut-off $x_0$ is given by $\sim \Lambda_\text{QCD}/P_z$. Notably, when $\xi < x_0$, LaMET cannot make reliable predictions for the ERBL region. Without exact information of the non-perturbative regions, we interpolate the results with different strategies to allow large variations representing the systematic uncertainties in these regions. Then we apply the resummed inverse matching kernel to the interpolated quasi-GPD and reproduce the original GPD model accurately in $\mathcal{X}$, despite the large difference in the interpolating function outside $\mathcal{X}$. The agreement suggests that our method is self-consistent, and also demonstrates that the inverse matching will not spread out the systematic uncertainties from non-perturbative effects and power corrections outside the region $\mathcal{X}$.

\begin{acknowledgments}
We thank Yushan Su and Zhite Yu for valuable discussions. 
The work of JH and HL is partially supported by the US National Science Foundation under grant PHY 1653405 ``CAREER: Constraining Parton Distribution Functions for New-Physics Searches'' and grant PHY~2209424, 
and by the Research Corporation for Science Advancement through the Cottrell Scholar Award.
The work of RZ and YZ is partially supported by the U.S. Department of Energy, Office of Science, Office of Nuclear Physics through Contract No.~DE-AC02-06CH11357, the Scientific Discovery through Advanced Computing (SciDAC) award \textit{Fundamental Nuclear Physics at the Exascale and Beyond}, and the Quark-Gluon Tomography (QGT) Topical Collaboration under contract no.~DE-SC0023646, and
Laboratory Directed Research and Development (LDRD) funding from Argonne National Laboratory, provided by the Director, Office of Science, of the U.S. Department of Energy under Contract No.~DE-AC02-06CH11357.

\end{acknowledgments}

\bibstyle{jhep}
\input{theory_paper_v2.bbl}

\end{document}

%% file: theory_paper_v2.bbl
\providecommand{\href}[2]{#2}\begingroup\raggedright\endgroup